\documentclass{elsarticle}

\usepackage{graphicx,psfrag,epsf}
\usepackage{enumerate}
\usepackage[utf8]{inputenc}
\usepackage{multicol}
\usepackage{multirow}
\usepackage{enumitem}
\usepackage{amsmath, amscd, amssymb, amsthm, xspace, graphicx}

\newtheorem{prop}{Proposition}
\usepackage[margin=1in]{geometry} 
\usepackage[all]{xy}
\usepackage{subcaption}
\usepackage{float}
\usepackage{mathtools} 
\usepackage{listings}
\usepackage{comment}
\usepackage{color}
\usepackage{orcidlink}
\usepackage[toc,page]{appendix}
\newcommand{\blind}{0}

\newcommand{\diffd}{\textnormal{d}}

\newcommand{\pderiv}[2]{\frac{\partial #1}{\partial #2}}
\newcommand{\inprod}[2]{\left\langle #1,#2\right\rangle}
\newcommand{\abs}[1]{\left| #1 \right|}

\newcommand{\norm}[1]{\left|\left|  #1 \right| \right|}
\newcommand{\LtwoStwo}{\mathcal{L}^2\left( \mathbb{S}^2 \right)}

\newcommand{\Stwo}{\mathbb{S}^2}
\newcommand{\mat}[1]{\mathbf{#1}}
\newcommand{\m}[1]{\mathbf{#1}}
\newcommand{\mq}{\m{Q}}
\newcommand{\md}{\m{D}}
\newcommand{\mQ}{\mq}
\newcommand{\mD}{\m{D}}
\newcommand{\mR}{\m{R}}
\newcommand{\Floor}[1]{\lfloor #1 \rfloor }
\newcommand{\Ceil}[1]{\lceil #1 \rceil }
\newcommand{\tr}[1]{\mathrm{tr}\left(#1\right)}
\newcommand\numberthis{\addtocounter{equation}{1}\tag{\theequation}}
\newcommand{\degrees}{^{\text{o}}}

\journal{Spatial Statistics}

\begin{document}

\begin{frontmatter}



\if0\blind
{
\title{Modeling Large Nonstationary Spatial Data with the Full-Scale Basis Graphical Lasso}
\author[1]{Matthew LeDuc}
\author[1]{William Kleiber}
\author[1,2]{Tomoko Matsuo}
\affiliation[1]{organization= {University of Colorado-Boulder Department of Applied Mathematics},
city={Boulder}, 
state={CO}, 
postcode={80309}}
\affiliation[2]{organization= {University of Colorado-Boulder Ann \& H.J. Smead Department of Aerospace Engineering Sciences},
city={Boulder}, 
state={CO}, 
postcode={80309}}
} \fi

\if1\blind
{
  \bigskip
  \bigskip
  \bigskip
  \begin{center}
    {\LARGE\bf  Modeling Large Nonstationary 
  Spatial Data with the
  Full-Scale Basis Graphical Lasso}
\end{center}
  \medskip
} \fi

\begin{abstract}
We propose a new approach for the modeling large datasets of nonstationary spatial processes that combines a latent low rank process and a sparse covariance model. 
The low rank component coefficients are endowed with a flexible graphical Gaussian Markov random field model.
The utilization of a low rank and compactly-supported covariance structure combines the full-scale approximation and the basis graphical lasso; we term this new approach the full-scale basis graphical lasso (FSBGL). 
Estimation employs a graphical lasso-penalized likelihood, which is optimized using a difference-of-convex scheme. 
We illustrate the proposed approach on synthetic fields as well as with a challenging high-resolution simulation dataset of the thermosphere. 
In a comparison against state-of-the-art spatial models, the FSBGL performs better at capturing salient features of the thermospheric temperature fields, even with limited available training data.
\end{abstract}

%



\end{frontmatter}





\section{Introduction}
\label{sec:intro}

Spatially-distributed datasets have become increasingly large and complicated in the past few decades. 
The field of spatial statistics has put tremendous effort into developing statistical models for such data, often by exploiting notions of sparsity or some rank-dependent structure in the covariance matrix. 
A typical statistical representation of a data process $Y(\vec{x})$ over domain $\vec{x}\in \Omega \subseteq {\mathbb R}^d$ takes the form 
\begin{align}
  \label{eq:obs.model}
  Y(\vec{x}) = \mu(\vec{x}) + Z(\vec{x}) + \varepsilon(\vec{x})
\end{align}
where $\mu(\vec{x})$ represents nonrandom mean fluctuation, $Z(\vec{x})$ represents mean zero spatially-correlated deviations and $\varepsilon(\vec{x})$ represents a mean zero generalized white noise process. 
Most efforts in the statistical literature focus on the developing flexible models for $Z(\vec{x})$.  
In this work we propose a new model for $Z(\vec{x})$ that synthesizes multiple recent proposals in the literature, giving rise to a flexible family of stochastic processes that can handle nonstationary processes and large observational datasets. 

The field of spatial statistics has embraced basis function expressions of $Z(\vec{x})$ as a useful, interpretable and flexible approach.
Endowing the basis functions with random coefficients allows for flexibility in handling different types of process behavior that can also allow for feasible estimation and inference schemes even with very large datasets. 
Some well-known examples include fixed rank kriging (\cite{frk}), LatticeKrig (\cite{nychka}), the stochastic partial differential equation (SPDE) approach (\cite{lind:11,inla,sigrist}), and the basis graphical lasso (\cite{bgl,mgl}). 
A comprehensive review of these methods is provided by \cite{annurev:/content/journals/10.1146/annurev-statistics-040120-020733}. 

In the late 2000s, fixed rank kriging (\cite{frk}) considered a specification of $Z(\vec{x})$ as a basis function expansion, with parameters estimated by least squares. 
Although convenient for large datasets, fixed rank kriging has been criticized for its inability to capture small scale dependence (\cite{steinlowrank}). 
This work inspired \cite{nychka} in which $Z(\vec{x})$ has a very large number of compactly supported functions in a multiresolution expansion which can even exceed the number of observations -- the key regularization idea was to enforce a spatial autoregressive (SAR) structure on the coefficient graphical model, thereby reducing the parameter space and affording computational feasibility using sparse matrix methods. This proposal, LatticeKrig, was shown to approximate standard stationary process covariance models under appropriate structuring of the SAR graphical component. 
A similar concept is applied in \cite{nngp}, where a nearest-neighbor structure in space rather than among the coefficients allows the dataset to be described using a directed acyclic graph. By approximating the full density at each point by conditioning on some number of points a reference set (for example, a randomly chosen subset of the spatial locations) instead of the entire dataset, the authors are able to preserve the proper rank of the covariance matrix while making parameter estimation feasible. Another similar concept is applied in \citep{HERMES2025100893} to the problem of modeling intercropping data with asymmetric effects across species, with the graphical model describing relationships between crop species rather than over space.

In \cite{bgl}, the LatticeKrig approach was extended to accommodate nonstationary Gaussian graphical models by imposing a graphical lasso regularization term in a likelihood-estimation framework. Their approach allowed for a nonstationary model that can be efficiently estimated for very large spatial datasets. This approach is termed the basis graphical lasso (BGL). 

When the basis expression contains a small number of functions (that is, small compared to the size of the dataset), it is helpful to include an additional stochastic process that represents small-scale behavior (\cite{steinlowrank}). This leads us to consider the representation of the process $Z(\vec{x})$ as 
\begin{align}  \label{eq:Z.Z1.Z2}    
  Z(\vec{x}) = Z_1(\vec{x}) + Z_2(\vec{x})
\end{align}
where both $Z_1$ and $Z_2$ are mean zero, spatially correlated processes.  
The key vision for $Z_1$ and $Z_2$ is that $Z_1$ aims to capture large-scale nonstationary structures via a low-rank basis expression, while $Z_2$ is designed for small-scale locally-correlated behavior which we assume to be stationary. 
Endowing $Z_2(\vec{x})$ with a compactly supported covariance structure gives rise to the so-called full scale approximation considered in \cite{steinfullscale} and \cite{sangfullscale}. 
This mixture gives rise to a flexible modeling framework that has a feasible estimation procedure for large datasets. This was expanded upon via a fused Gaussian process model in \cite{ma.kang}, where the process $Z_2(\vec{x})$ is broken up into $M$ disjoint regions described by independent conditional autoregressions that are then fit to the data. This makes fitting the model much more manageable, as instead of fitting to an entire large dataset, which even with a sparse model can be unwieldy, the fitting is done on $M$ smaller datasets.  

Further extensions were considered in \cite{katzfuss2013},\cite{katzfuss2017}, and \cite{katzfuss2020} with the M-RA, where instead of a single low-rank process the domain is divided up via a recursive domain partitioning scheme or a set of multiresolution tapering functions. This model takes the full-scale approximation of \cite{sangfullscale} and effectively iteratively applies it, so that the full-scale approximation can be seen as the M-RA with one level of resolution (\cite{katzfuss2017}). However, unlike the full-scale approximation, which only rely on a covariance model for the small-scale process and uses empirical orthognal functions (EOFs) derived from the sample covariance for the low-rank portion, this method relies on a covariance model for the entire process, which is then used to determine the basis functions. Additionally, processes on disjoint domains are independent in this construction, so domains must be carefully chosen to respect the assumed covariance length scales. 

In this work, we combine the approaches of \cite{bgl} and \cite{sangfullscale} to yield a flexible and interpretable approach to handling large observational datasets of nonstationary spatial processes. 
Our proposal encapsulates ideas from fixed rank kriging, the full scale approximation, and the basis graphical lasso within one single framework that both allows for flexible nonstationary covariance structures that can also capture small scale correlation structure. For this reason, we call our model the \textit{full-scale basis graphical lasso} (FSBGL).

To demonstrate the efficacy of this model we assess its ability to represent the statistical structure of lower-thermospheric neutral temperature fields simulated by a large Earth system atomosphere model (\cite{liu.gravity.waves}). The temperature in the lower thermosphere is highly spatially structured and temporally variable due to the influence of atmospheric waves originating from meteorological weather events. Thermospheric gases exert considerable aerodynamic forces on space vehicles and objects, and the spatial structures of lower thermosphere temperature fields considerably affect their reentry trajectory due to their relationship to the neutral mass density (\cite{leonardetal}).  The model developed in this study can help engineers, who usually rely on standard atmospheric gas models, account for the effects of spatially structured mass density fields on the expected forces experienced by vehicles during reentry.

\section{Methods}

In this section we describe the general modeling framework and estimation procedure with particular attention to the case when the number of observation locations is large.

\subsection{Statistical Model}

We consider a statistical model of the form given by \eqref{eq:obs.model} and \eqref{eq:Z.Z1.Z2}, with the additional assumption that
\begin{align}  \label{eq:Z.frk}
  Z_1(\vec{x}) = \sum_{j=0}^{J} \beta_{j} \psi_{j}(\vec{x})
\end{align}
for a set of spanning functions $\{\psi_{j}(\vec{x})\}_{j=0}^{\infty} \subset \mathcal{L}^2(\Omega)$ and where the coefficients $\beta_j$ are jointly distributed according to a mean zero multivariate Gaussian graphical model. 
Specifically, we will let $\vec{\beta} = (\beta_0,\beta_1,\ldots,\beta_J)^T$ where $\beta\sim N(0,\mat{\mQ}^{-1})$ with $0$ a mean vector of zeros of length $J+1$ and $\mat{Q}$ the precision matrix. 
It is important to note that we assume $\mat{Q}$ is sparse, which will be enforced through regularization in our proposed estimation scheme. 
Next, let $Z_2$ be an isotropic Gaussian process with compactly supported isotropic covariance function $C_2(\| \vec{h} \|) = {\rm Cov}(Z_2(\vec{x}+\vec{h}),Z_2(\vec{x}))$.
Finally, let $\varepsilon$ be a mean zero Gaussian white noise process with variance $\tau^2$.
The major obstacles given this model are in specifying the structure of the graphical model corresponding to $\mat{Q}$ and estimating parameters of the covariance function $C_2$.  
Synthesizing the assumptions in (\ref{eq:obs.model}), (\ref{eq:Z.Z1.Z2}) and (\ref{eq:Z.frk}), our full proposed model for the observation process is 
\begin{align}
  \label{eq:full.obs.model}
  Y(\vec{x}) &= \mu(\vec{x}) + \sum_{j=0}^{J} \beta_{j} \psi_{j}(\vec{x}) + Z_2(\vec{x}) + \varepsilon(\vec{x}).
\end{align}

\subsection{Estimation}

It is helpful to revisit how extant approaches handle estimation of spatial parameters to contrast these with the FSBGL. 
Suppose we have a set of $N$ independent realizations $Y_i(\vec{x})$ of the process from \eqref{eq:full.obs.model} at locations $\vec{x}_1,\ldots,\vec{x}_n\in\Omega$, setting $\vec{Y}_i=(Y_i(\vec{x}_1),\ldots,Y_i(\vec{x}_n))^T$ and then $\mat{Y} = (\vec{Y}_1,\vec{Y}_2,...,\vec{Y}_N)$. 
For now, assume the mean function $\mu(\vec{x}) = 0$, but in general it will have the form of a regression whose coefficients can be estimated by generalized least squares. 
Especially, consider the case when $n$ is moderate-to-large (tens of thousands or more). 
The negative log-likelihood is given by 
\begin{equation}\label{eq:loglike}
    L(\mat{Q},\mat{D}|\mat{Y}) =  \log \det\left(\mat{\m{\Psi}} \mat{Q}^{-1} \mat{\m{\Psi}}^T+\mat{D} \right) + \\  \text{tr}\left(\mat{S}\left(\mat{\m{\Psi} Q}^{-1} \mat{\m{\Psi}}^T+\mat{D} \right)^{-1}\right)
\end{equation}
where $\mat{D}$ captures the compactly supported portion of the covariance as well as the white noise, $\mat{S}$ is the sample covariance matrix of the data $\m{Y}$, and $\mat{\m{\Psi}}$ is a matrix such that $\mat{\m{\Psi}}_{ij} =\psi_j(\vec{x}_i)$. 
The maximum-likelihood estimates of $\m{Q}$ and $\m{D}$ are the matrices $\Tilde{\m{Q}}$ and $\Tilde{\m{D}}$ that satisfy $\m{S} = \m{\m{\Psi}} \Tilde{\m{Q}}^{-1} \m{\m{\Psi}}^T+\Tilde{\m{D}} $, however these matrices are difficult to estimate, dense, and often are not uniquely determined by the data. 

In the fixed-rank kriging scheme this is handled by setting $Z_2=0$, and then $\tau^2$ and the covariance matrix of $\vec{\beta}$ is estimated via least-squares (\cite{frk}).
If $\m{\m{\Psi}}$ has a QR decomposition, then \cite{frk} provides closed-form solutions.
In contrast, LatticeKrig solves this via parameterization of $\m{Q}$ as a block-diagonal set of stationary Gaussian Markov random field precision matrices, specified as spatial autoregressions.
In LatticeKrig the basis is fixed to be Wendland Radial Basis Functions and $\m{Q}$ is a paraeterized spatial autoregression with independent levels of resolution.
The exact computational strategy is outlined in \cite{nychka}.  
Analogous to fixed rank kriging, within our modeling framework this is captured by setting $Z_2(\vec{x})$ identically to zero.

One drawback of the LatticeKrig approach is that the stochastic coefficients are restricted to (approximate) stationarity using the spatial autoregressive GMRF.  
The basis graphical lasso (BGL) relaxes the stationarity assumption, allowing for $\m{Q}$ to represent a GMRF structure through a sparsity pattern that is estimated by applying a graphical lasso penalty to the likelihood (\cite{bgl}). 
This is equivalent to attempting to recover a graphical model for the conditional dependencies among the coefficients, and it means that the whole (potentially dense) coefficient covariance matrix need not be directly estimated. 
Assuming that $\m{D }= \tau^2 \m{I}$, the BGL minimizes the penalized log-likelihood 
\begin{equation}
    \label{eq:bgllike}
     L(\m{Q},\tau^2|\m{Y}) =  \log \det\left(\m{\m{\Psi} Q}^{-1} \m{\m{\Psi}}^T+\tau^2\m{I} \right) + \\  \text{tr}\left(\m{S}\left(\m{\m{\Psi} Q}^{-1} \m{\m{\Psi}}^T+\tau^2\m{I} \right)^{-1}\right) +||\m{\Lambda} \circ \m{Q}||_1
\end{equation}
where $\norm{\m{\Lambda} \circ \m{Q}}_1=\sum_{i,j}\abs{\m{\Lambda}_{ij}\m{Q}_{ij}}$.

In Proposition 1 in \cite{bgl} it was shown that if $\tau^2$ is known, minimizing the log-likelihood in \eqref{eq:bgllike} is equivalent to minimizing the log-likelihood given by

\begin{align}
  \begin{split}\label{eq:bglprop1}
    L(\m{Q} | \m{Y},\tau^2) &=  \log \det\left(\m{Q}+\tau^{-2} \m{\m{\Psi}^T\m{\Psi}}\right) - \log \det\left(\m{Q}\right) \\
    &- \textnormal{tr}\left(\tau^{-4} \m{\m{\Psi}^T S \m{\Psi}}\left(\m{Q}+\tau^{-2} \m{\m{\Psi}}^T\m{\m{\Psi}}\right)^{-1}\right) + \norm{\m{\Lambda} \circ \m{Q}}_1
  \end{split}
\end{align}
which can be minimized via a difference of convex scheme outlined therein. There, $\tau^2$ was calculated with the assumption that $\m{Q}=\alpha \m{I}$, and then this value was used in the minimization of \eqref{eq:bglprop1}. This model was shown to give superior predictions to other state of the art models such as LatticeKrig and fixed-rank kriging when applied to various climatological datasets. It was also shown to be able to accurately recover the structure of a known precision matrix (\cite{bgl}). 
However, assuming $\m{D}=\tau^2\m{I}$ is again equivalent to setting $Z_2$ to identically zero in our setup. 

In \cite{steinfullscale} and \cite{sangfullscale}, two similar models are proposed for large spatial datasets. These assume that the field has the form given by \eqref{eq:Z.Z1.Z2}, where the process $Z(\vec{x}) = Z_1(\vec{x})+Z_2(\vec{x})$ has a low-rank and small-scale portion. These sorts of models have been called full-scale approximations based on terminology from \cite{sangfullscale}. This leads to the covariance structure
\begin{equation}
    \label{eq:fullscalecov}
    \m{\m{\Psi}\Sigma \m{\Psi}}^T +\m{D}
\end{equation}
where $\m{\m{\Psi}}$ is a matrix of basis functions, $\m{\Sigma}$ the covariance matrix of the coefficients, and $\m{D}$ the covariance of $Z_2$. 

A key insight in the full-scale approximation is to endow $Z_2$ with a compactly supported covariance function, making $\m{D}$ sparse. 
Then the likelihood \eqref{eq:loglike} can be rewritten using the Woodbury formula and properties of the determinant, and parameters of $\m{D}$ are estimated from this using MCMC. In \cite{steinfullscale}, the basis functions were taken to be Legendre polynomials, and the Cholesky factor of $\m{\Sigma}$ was estimated after the parameters of $\m{D}$. This was done with a small number of basis functions, keeping estimation simple. As noted in \cite{steinfullscale}, estimating a dense Cholesky factorization is $O(n^3)$, meaning that even though this allows for non-trivial relationships between the basis functions, the estimation procedure is unwieldy for large sets of basis functions without further modification.
In \cite{sangfullscale}, this was simplified by using empirical orthogonal functions to represent the low rank portion of the model. This keeps $\m{\Sigma}$ a diagonal matrix that is estimated simultaneously with $\m{\m{\Psi}}$. 
Unfortunately neither of these approaches work well for a large number of basis functions as the covariance matrix $\m{\Sigma}$ would require some regularization, or severe limiting assumptions (e.g., being diagonal as in \cite{sangfullscale}). 
In this work we relax these restrictions by incorporating lasso-like regularization on the inverse covariance matrix of the stochastic coefficients.

\subsubsection{Estimation for the Full-Scale Basis Graphical Lasso (FSBGL)}
 
Both the full-scale approximations and the BGL rely on simplification of the covariance structure, either by keeping $\m{Q}$ diagonal via orthonormal basis functions such as in \cite{sangfullscale}, keeping the number of basis functions extremely small to allow for estimation of the Cholesky decomposition of the covariance as in \cite{steinfullscale}, or by assuming that $Z_2(\vec{x})$ is identically zero.
Our proposal is to combine both methods by allowing for structure in both $\mq$ and $Z_2$ while still maintaining computational feasibility. 
 
If, instead of letting $\md=\tau^2 \m{I}$, we allow $\md$ to be an arbitrary positive-definite matrix, then the regularized likelihood is given in the following proposition, a special case of which appeared in \cite{mgl}.
\begin{prop}\label{thm:prop1}
    Let $\md$ be an arbitrary covariance matrix, then the regularized likelihood in \eqref{eq:bgllike} is equivalent to 
    \begin{equation}\label{eq:prop1}
    \begin{split}
        L(\mQ|\m{Y},\mD) &= \log \det\left(\mQ+ \m{\m{\Psi}^T D}^{-1} \m{\m{\Psi}}\right) - \log \det\left(\mQ\right) \\
         &- \textnormal{tr}\left(\m{\m{\Psi}}^T \mD^{-1}\m{S} \mD^{-1} \m{\m{\Psi}}\left(\mQ+\m{\m{\Psi}}^T \mD^{-1} \m{\m{\Psi}}\right)^{-1}\right) + ||\m{\Lambda} \circ \mQ||_1
        \end{split}
    \end{equation}
\end{prop}
The proof is presented in Appendix A, but it is a straightforward generalization of the proof presented in \cite{bgl}. 
This gives the field covariance the form of a low-rank representation plus a high-resolution component represented by $\mD$, which is similar to the full-scale approximations discussed in \cite{steinfullscale} and \cite{sangfullscale}. 
The matrix solves in (\ref{eq:prop1}) can be made efficient by endowing $Z_2$ with a compactly supported covariance function, imposing sparsity in $\mD$.
For this reason, we call this model the full-scale basis graphical lasso (FSBGL).

As in \cite{bgl}, the main motivation for using Eq. \eqref{eq:prop1} rather than Eq. \eqref{eq:loglike} to fit the model is computational complexity.
If we are to use Eq. \eqref{eq:loglike} to fit the model,at each step we must compute the Cholesky decomposition of the matrix $\mat{\m{\Psi}} \mQ^{-1}\mat{\m{\Psi}}^T+\mD$, which is an $O(n^3)$ operation that is not helped by the fact that this matrix will, in general, be dense. Then once the Cholesky factor $\mat{L}$ is calculated, the log-determinant is calculated in $O(n)$ and the trace can be calculated as $\frac{1}{n}\| \mat{L}^{-1}\mat{Y}\|_{F}^2$, which is $O(n^2N+n^2)$.
Additionally, the calculation of $\mat{\m{\Psi}} \mQ^{-1}\mat{\m{\Psi}} $ is $O(n^2J + J^2n +J^3)$, leading to overall complexity that is $O(n^3)$.

In our preferred implementation, Eq. \eqref{eq:prop1}, we are able to leverage the sparsity of $\mD$ to improve the situation. First, calculating the Cholesky decomposition of $\mD=\mat{L}_D\mat{L}_D^T$ is an $O(n^3)$ operation naively, however since it is known it only needs to be done once, and the sparsity of $\mD$ will accelerate both this decomposition and the resulting applications of $\mat{L}_D$ greatly compared to the previous case where all involved matrices were dense. In the event that $\mD$ is a banded matrix, for example, this can be done in $O(k^2n)$ time, where $k$ is the number of non-zero diagonals. If $\mD$ is block-diagonal this will carry into the Cholesky decomposition, increasing the speed of those calculations as well. Then, the factors $\mat{\m{\Psi}}^T\mat{D}^{-1}\m{\m{\Psi}}$ and $\m{\m{\Psi}^TD}^{-1}\m{SD}^{-1}\m{\m{\Psi}}$ can be precomputed as well, and in fact since $tr(\m{A}^T\m{A})=\|\m{A}\|_F^2$ the trace term can be made even simpler to compute, requiring $O(J^3n^2)$ operations per step of the algorithm and only requiring that $\m{\m{\Psi}}^T\mD^{-1}\m{Y}$ be precomputed rather than $\m{\m{\Psi}}^T\mD^{-1}\m{SD}^{-1}\m{\m{\Psi}}$. The remaining operations are at worst $O(J^3)$, leading to significant computational improvements.

As suggested by \cite{bgl}, we can write the likelihood as a sum of convex and concave parts, and we can solve the optimization for $\mQ$ by a Difference of Convex (DC) scheme. 
The convex part is given by $-\log \det(\mQ) + ||\m{\Lambda} \circ \mQ||_1$ and the concave part by $\log \det\left(\mQ+ \m{\m{\Psi}}^T \mD^{-1} \m{\m{\Psi}}\right) - \textnormal{tr}\left(\m{\m{\Psi}}^T \mD^{-1}\m{S} \mD^{-1} \m{\m{\Psi}}\left(\mQ+\m{\m{\Psi}}^T \mD^{-1} \m{\m{\Psi}}\right)^{-1}\right) $, and so the DC scheme for minimization is given by
\begin{align*}
        \hat{\mQ}_{j+1}|\hat{\mQ}_j,\mD &=\underset{\mQ>0}{\text{argmin}} \left[-\log\det(\mQ)+\text{tr}(\nabla g(\hat{\mQ}_j)\mQ)+||\m{\Lambda} \circ \mQ||_1\right] \\
        \nabla g(\mQ_j) &=(\mQ_j+\m{\m{\Psi}}^T \mD^{-1}\m{\m{\Psi}})^{-1}(\mQ_j+\m{\m{\Psi}}^T\mD^{-1}\m{\m{\Psi}}+  \m{\m{\Psi}}^T\mD^{-1}\m{SD}^{-1}\m{\m{\Psi}})(\mQ_j+\m{\m{\Psi}}^T\mD^{-1}\m{\m{\Psi}})^{-1} \numberthis \label{eq:dcupdate}
\end{align*}
The derivation is provided in Appendix B, and since the same terms can be precalculated the algorithm achieves similar computational complexity. 
Successive calculations of $\hat{\mQ}_{j+1}$ are performed using QUIC (\cite{JMLR:v15:hsieh14a}). 
We say that the method has converged when, for a given $\delta$, $\frac{\norm{\hat{\mQ}_{j+1}-\hat{\mQ}_j}_{F}}{\norm{\hat{\mQ}_j}_{F}} < \delta$. In this paper, we set $\delta=0.02$.

To fit the model, we need to first produce an estimate of $\mD$ to be used in the minimization in \eqref{eq:dcupdate}. To do this, we adapted the algorithm used in \cite{bgl} to fit the nugget variance $\tau^2$. We make the assumption that $\mQ = \alpha \m{I}$, and then jointly fit $\alpha$ and $\mD$. 
We prefer to estimate this way because there is not a clear path to estimating a full model for $\mQ$ and $\mD$ simultaneously. Once we have determined $\mD$, we can then use Eq. \eqref{eq:dcupdate} to get the final estimate of $\mQ$.
Experiments on simulated data showed that re-estimating $\mD$ after estimating a full $\mQ$ did not lead to substantial improvement in the performance of the model, changing the updated likelihood by less than $0.4 \%$ on average. Similar behavior was also seen in \cite{bgl}. 

By assigning a covariance model to $Z_2(\vec{x})$ with parameters $p$, we can determine the best fit $\hat{p}$ by minimizing the likelihood in (\ref{eq:prop1}). Along with the assumption that $\mQ=\alpha \m{I}$, this leads to the optimization problem

\begin{equation}\label{eq:invertD}
    \begin{split}
    \hat{p},\hat{\alpha} = \underset{p,\alpha}{\text{argmin}} \log\det(\alpha \m{I}+\m{\m{\Psi}}^T\mD(p)^{-1}\m{\m{\Psi}})+ \log\det(\mD(p)) - \log\det(\alpha \m{I}) - \\ 
    \text{tr}(\m{\m{\Psi}}^T\mD(p)^{-1}\m{SD}(p)^{-1}\m{\m{\Psi}}(\alpha \m{I}+\m{\m{\Psi}}^T\mD(p)^{-1}\m{\m{\Psi}})^{-1}) + \text{tr}(\m{SD}(p)^{-1}) 
    \end{split}
\end{equation}
which can be derived as an intermediate step in the derivation of \eqref{eq:prop1}.
This can be fit by any standard optimization method, although care should be taken to make sure that operations on $\mD$ do not become prohibitively expensive. One option for this would be to parameterize $\mD$ by a Markov Random Field or other method that allows direct parameterization of the inverse, another might be to give $\mD$ block-diagonal structure or directly parameterize the Cholesky decomposition, as done in \cite{MUSCHINSKI2024261}, although if one is not careful the latter could greatly increase the number of parameters to estimate in return. We also recommend applying a derivative-free optimization method to avoid the need to calculate finite-difference approximations to the gradient, which in general will not have a closed form, thus avoiding extra likelihood evaluations. In the event that $n$ is small and there are a small number of parameters, this is less of a concern.

\subsubsection{Model Selection}

Once $\hat{p}$ is known, the precision matrix $\mQ$ is fit via the optimization problem in Equation \eqref{eq:dcupdate}, with the penalty matrix given by 
\begin{equation*}
  \m{\Lambda}_{ij}=
  \begin{cases}
    \lambda & i\ne j \\
    0 & i=j.
  \end{cases}
\end{equation*}
Not penalizing the diagonal of the precision matrix means that the estimated marginal precisions of the coefficients will be the maximum likelihood estimators. 

Since we fit the model for a range of $\lambda$, we need a model selection criterion to determine the optimal value of $\lambda$ for each model. In order to avoid overfitting, we choose to use the conditional Akaike information criterion in the form (\cite{hatmatrixAIC})
\begin{equation}\label{eq:hatmarixaic}
    cAIC(\lambda) = -2 \ell( \mQ,\mD|\m{Y} ) + 2 df_{E}
\end{equation}
where $\ell(\mQ,\mD|\m{Y})$ is the log-likelihood and $df_{E}$ is the effective degrees of freedom, given by $df_E=\tr{\hat{\m{H}}}+|p|$. Here, $|p|$ is the number of parameters in the model for $\mD$ and $\hat{\m{H}}$ is the hat matrix, given by 
\begin{equation}
    \hat{\m{H}} =  \m{\m{\Psi}}\left(\m{\m{\Psi}}^T\mD^{-1}\m{\Psi}+\mQ\right)^{-1}\m{\Psi}^T\mD^{-1}
\end{equation}
The hat matrix is the optimal predictor of the field given the data, and its trace, which lies between $0$ and $J+1$, is a measure of the influence of $\mQ$. If the trace is small, that means that $\mQ$ is large, i.e. the estimates of $\beta$ have small variance, and if the trace is large, then our estimates of $\beta$ have large variance. For this reason, keeping $\tr{\hat{\m{H}}}$ small prevents overfitting. 

\subsubsection{Implications of Misspecification in $\mD$}
We conclude the Methods section with a brief discussion of how misspecification in the small-scale covariance $\mD$ affects estimation of $\mQ$. Although the processes $Z_1(\vec{x})$ and $Z_2(\vec{x})$ are assumed independent, estimation of $\mQ$ depends on $\mD$ through the joint covariance structure in Eq. \eqref{eq:prop1}. When the true small-scale covariance is $\mD_*$ but a misspecified covariance $\mD = \mD_* + h\mR$ is used in its place, the estimation of $\mQ$ will be most strongly influenced by the portion of $\m{R}$ that is represented within the low-rank portion, similar to the spatial confounding seen in hierarchical modeling (e.g. \cite{hanks2015restricted}). This changes how variance is spread between the different parts of the model and may bias the estimate of $\mQ$.
Depending on the magnitude of this effect, such misspecification can also affect the inferred graphical structure of $\mQ$ through the lasso penalty, leading to spurious removal or introduction of edges. 

\section{Simulation Studies}\label{sec:simdata}

In this section we consider a set of simulation studies to investigate whether the proposed FSBGL is able to recover the structure in $\mQ$, as well as accurately determine the parameters of the stationary residual process. 
We simulate random fields $Z(\vec{x})$ on $[0,1]^2$ of the form 
\begin{equation}\label{eq:simstudy_datafn}
    Z(\vec{x}) = \sum_{i=0}^J \beta_i\psi_i(\vec{x}) +Z_2(\vec{x})+\epsilon(\vec{x})
\end{equation}
using basis functions $\psi(\vec{x})=\frac{1}{2}\cos\left(2\pi f_x x\right)\cos(2\pi f_y y)$ for $f_x,f_y=0,1,..,10$, leading to $J= 121$ total basis functions. The field is simulated at $n=1681$ evenly spaced locations. 

\begin{figure}[t]
    \centering
    \includegraphics[width=\linewidth]{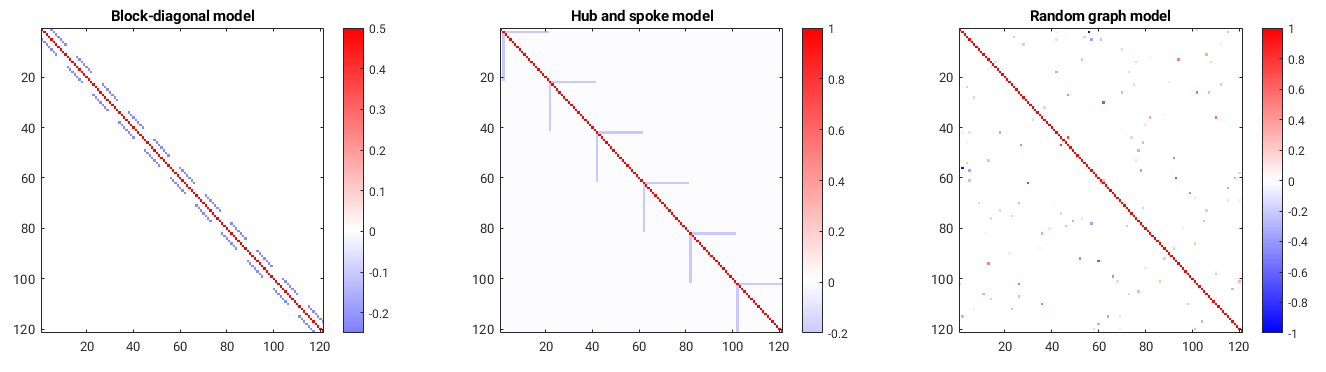}
    \caption{Three different models of $\mQ$ used to test the ability of the FSBGL to recover structure in a known precision matrix in the simulation studies.}
    \label{fig:models_of_Q}
\end{figure}
We consider multiple common types of precision matrix structure for the coefficients $\vec{\beta}$ in Eq. \eqref{eq:simstudy_datafn} and use a tapered Mat\'ern covariance model for $Z_2(\vec{x})$ (\cite{furrercovtapering}). 
The small-scale variation covariance is given by 
\begin{equation}
    C(d) = C_{\nu}\left(\frac{d}{r}\right)C_{\theta}(d)
\end{equation}
where $C_{\theta}(d)$ is the tapering function and $C_{\nu}\left(\frac{d}{r}\right)$ is a Mat\'ern covariance with smoothness parameter $\nu$ and range parameter $r$,
\begin{equation}
    C_{\nu}\left(\frac{d}{r}\right) = \frac{2^{1-\nu}}{\Gamma(\nu)}\left(\frac{d}{r}\right)^\nu K_{\nu}\left(\frac{d}{r}\right)
\end{equation}
where $K_{\nu}$ is a modified Bessel function of the second kind. 
Fields with a Mat\'ern correlation have $\Floor{\nu}$ mean-square derivatives \citep{steinbook}. The taper function is a Wendland radial basis function \citep{wendlandrbf}. We fit every parameter of the tapered Mat\'ern model, along with a process variance and the logarithm of the nugget effect variance $\tau^2$, totaling 5 parameters. The simulation setup is described in Table \ref{tab:paramvals_simstudy}.

\begin{table}[]
    \centering
    \begin{tabular}{|c|c|c|}
    \hline
         Parameter& Description & Truth\\
         \hline
         $N$&Number of independent field replicates&10,30,50,100\\
         $n$&Number of spatial locations&1681\\
         $\psi_i(\vec{x})$&Basis functions & Cosine series \\
          $J$&Number of basis functions&121 \\
         $\mQ$&  Precision matrix of $\vec{\beta}$ &Varying (Fig. \ref{fig:models_of_Q})\\
         $\sigma^2$&Variance of $Z_2$&1\\
         $\tau^2$&Nugget variance&0.01\\
         $\theta$&Wendland taper radius&$0.3$\\
         $r$&Mat\'ern correlation length&0.15\\
         $\nu$&Mat\'ern smoothness &0.5\\
         \hline
    \end{tabular}
    \caption{Experimental setup for the simulation study. Definition of experimental parameters as well as ground truth.}
    \label{tab:paramvals_simstudy}
\end{table}

Using the parameter values in Table \ref{tab:paramvals_simstudy}, we simulate $N=10,30,50,100$ independent realizations of the random field and report the median errors over 50 trials of each in the recovery of $\mQ$ and the parameters of the residual model in \ref{sec:simdata_appendix}. The focus is on three different models for $\mQ$, which we call the block-diagonal, hub and spoke, and random graph models. These models are shown in Figure \ref{fig:models_of_Q}. 
The results for the block diagonal model (on the left of Figure \ref{fig:models_of_Q}) are presented in Table \ref{tab:simresults_bdiag_inbody}. For the errors in the parameters we present the median absolute error and for the errors in $\mQ$ we calculate the median relative error in the Frobenius norm $\|\hat{\m{Q}}-\m{Q}\|_F/\|\m{Q}\|_F$, and likewise for $\mQ^{-1}$. In general we find that the model is able to recover the parameters of $Z_2$ with reasonable accuracy regardless of the number of samples, but the recovery of $\mQ$ is dependent on the number of samples $N$. 

\begin{table}[h!]
    \centering
    \begin{tabular}{|c|c|c|c|c|c|}
    \hline
        Parameter&Truth & N=10 &N=30  &N=50& N=100  \\
          \hline
    $\mQ$ & block-diag & 0.610 &0.444 &0.436 & 0.277\\      
    \hline
     $\mQ^{-1}$ & block-diag & 0.654 &0.515 &0.421 & 0.435\\      
    \hline
    Missed non-zero (\%) &N/A& 51.6&13.8&8.73&0 \\
    \hline
    Missed zero (\%) &N/A& 0.25&4.8&1.54&19.2 \\
    \hline
    $\sigma^2$&1&0.041 & 0.041&0.059 &0.041\\
    \hline
    $\log_{10}(\tau^2)$&-2& 0.224&0.184& 0.232&0.236\\
    \hline
    $\theta$&0.3&0.027& 0.019& 0.043& 0.035\\
    \hline
    $r$& 0.15&0.027& 0.043& 0.036& 0.044\\
    \hline
    $\nu$& 0.5&0.043& 0.073& 0.062& 0.060\\
    \hline
    $L(\hat{\mQ},\hat{p})/L(\mQ,p)$ &1&1.1&1.1&1.04&1.1\\
    \hline
    \end{tabular}
    \caption{Median error in recovery of residual parameters for varying number of independent replicates of the field using a block-diagonal model for $\mQ$, along with true values. All calculations are the median over 50 trials for a given number of independent replicates of the field. The error in the recovery of $\mQ$ is given by $\|\hat{\mQ}-\mQ\|_F/\|\mQ\|_F$ with $\|\cdot\|$ the Frobenius norm, and the error for the other parameters is the absolute error.}
    \label{tab:simresults_bdiag_inbody}
\end{table}

For the purposes of this study the errors for $\mQ$ and $\mQ^{-1}$ are calculated for $\lambda=10^k$ for $k=-1,-0.5,0,0.5,1$ and the reported values are the errors for the value of $\lambda$ that minimizes the error in $\mQ$. The optimal value of $\lambda$ decreases with the number of samples, as the sparsity penalty becomes less important with greater amounts of data. For small numbers of samples, then, the estimated $\mQ$ is typically diagonal, with additional structure recovered as the number of samples increases. The error in $\mQ$ is comparable to that achieved by the BGL in Table 1 of \cite{bgl}. 
\ref{sec:simdata_appendix} contains results for the other precision matrix structures, whose results are analogous to the block-diagonal case, as well as the full results for the investigated values of $\lambda$. These results show that, while the best penalty value in terms of recovering all zero entries is large and the best value in terms of recovering non-zero entries is small, as expected, the best overall values of $\lambda$ in terms of error in $\mQ$ and recovery of the sparsity pattern tend to be the same.
We also include visualizations of the estimated precision matrices, which suggest that the general precision structure is well-recovered once a few dozen realizations are available.

\section{Application to WACCM-X Neutral Temperature Fields}

In this section we illustrate the FSBGL and other competing models in a task to efficiently represent the multi-scale and nonstationary properties of thermospheric temperature simulated by a large numerical Earth system model. 
We compare both in-sample and out-of-sample model behavior, and show that the FSBGL provides a better representation of the process than state-of-the-art alternatives. 

\subsection{Data}

The National Center for Atmospheric Research (NCAR) Whole Atmosphere Community Climate Model with thermosphere and ionosphere extension (WACCM-X) is a high-top Earth system atmosphere model that extends from the surface to low-Earth orbit (LEO) altitudes ($\sim$500 km) (\cite{waccmx}). Recently, WACCM-X was run at a high-resolution configuration to model intricate structures of the thermosphere influenced by atomospheric wave forcing due to highly variable tropospheric weather (\cite{liu.gravity.waves}). Variability of the lower thermospheric temperature structures (80-150km) is not well represented in the standard atmosphere models, but it is critical to modeling aerodynamic forces experienced by vehicles and objects during reentry (\cite{leonardetal}). Thermospheric temperature fields derived from the data used in \cite{liu.gravity.waves} are used for illustrations of the method.

\begin{figure}
     \centering
     \includegraphics[width=\textwidth]{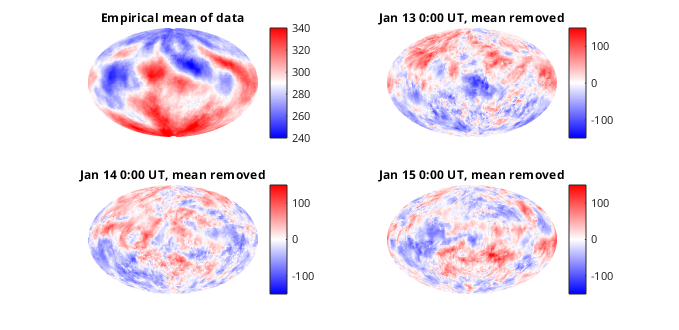}
     \caption{The sample mean of the neutral temperature data (top left) and 3 model runs from the same UT with the sample mean removed (others). The fields are plotted with a Hammer projection and units are Kelvin.}\label{fig:glowexample}
\end{figure}

The data we examine consist of 41 replicate samples of neutral temperature fields at a pressure level corresponding roughly to 130-135 km altitude. 
The data correspond to simulation outputs with a 6 hour time resolution from 0:00 UT on January 13 to 24 UT January 22 under solar minimum conditions. These data are available on a $0.25\degrees \times 0.25\degrees$ grid, but are downsampled to $2\degrees\times 2\degrees$ to alleviate computational concerns.  This also allows us to test the model's ability to predict spatial scales smaller than those contained in the training data, which are referred to as subgrid in this study.  
The model is fit to data from January 13 to January 20 (32 samples) and tested on the remaining data (9 samples).
The spatial coordinate system is rotated with the motion of the Sun to focus on spatial variability that is synchronized with solar heating. The center of the figures shown in this paper thus corresponds to local noon, and the edges to local midnight.
The empirical mean of these data, as well as three fields with the mean subtracted, computed on the $2\degrees\times 2\degrees$ grid, are shown in Figure \ref{fig:glowexample}. Additionally, the data generally show little autocorrelation between samples, as shown in \ref{sec:autocorr_appendix}. 
Dynamics in this altitude region is known to be dominated by the day-to-day variability of atmospheric waves originating from tropospheric weather. This allows us to treat each spatial field as independent across time in our analysis.

The dimension of these data, containing over 16000 spatial locations for each field, and the complicated structure seen in Figure \ref{fig:glowexample} challenge existing modeling frameworks. In the next subsection we detail the FSBGL modeling choices and in-sample model fit, followed by a detailed comparison against competing alternatives.

\subsection{FSBGL Specification}

The FSBGL of Equation (\ref{eq:full.obs.model}) requires specification of the mean function, the low rank component $Z_1(\vec{x})$, and the spatially-correlated small scale variation process $Z_2(\vec{x})$. 
We specify the mean function as an empirical mean of the values on the grid over each realization of the field.

\subsubsection{$Z_1(\vec{x})$}


For the low-rank portion of the model, we choose to write the field in a needlet expansion. 
Needlets are a multiresolution frame for data on $\Stwo$ first introduced in \cite{needletintro}, and additional details of their construction are given in \cite{cmbneeds} and \cite{RFSphere}. 
Letting $Y_{lm}(\vec{x})$ be the spherical harmonics, the needlet $\psi_{jk}(\vec{x})$ is given by 
\begin{equation}\label{eq:needletdef}
    \psi_{jk}(\vec{x}) = \sqrt{\lambda_{jk}}\sum_{l=\Ceil{B^{j-1}}}^{\Floor{B^{j+1}}}b\left(\frac{l}{B^{j}}\right)\sum_{m=-l}^{l}Y_{lm}(\xi_{jk})\overline{Y_{lm}(\vec{x})}.
\end{equation} 
Here, $j$ is a spatial resolution parameter, $k$ controls location, $B$ controls the extent of the needlets in harmonic space (and here will be set to 2), the weights and points $\{\lambda_{jk},\xi_{jk}\}$ form an exact cubature rule on $\mathbb{S}^2$ for polynomials of degree less than $2B^{j+1}$, and $b(t)$ is a non-negative $C^{\infty}$ function with compact support on the interval $[B^{-1},B]$ such that $\forall \eta >1, \sum_jb(\frac{\eta}{B^j})^2=1$. 
For such $b(t)$, the needlets are compactly supported in the harmonic domain and additionally decay faster than any rational function away from $\xi_{jk}$, meaning that they have excellent localization properties even though they do not have local support (\cite{RFSphere}).

In addition to providing localization in physical and harmonic space, they have several other important analytic properties. Along with the first spherical harmonic $Y_{00} = \frac{1}{\sqrt{4\pi}} = \psi_{-1,0}$ they form a Parseval-tight frame of $\LtwoStwo$ (\cite{needletintro}). 
This means that for all $f(\vec{x}) \in \LtwoStwo$, 
\begin{align*}
    f(\vec{x})&=\sum_{j,k}\beta_{jk}\psi_{jk}(\vec{x})\\
    \norm{f}_2^2 &= \sum_{j,k}\abs{\beta_{jk}}^2 \numberthis \\
    \beta_{jk} &= \int_{\Stwo} f(\vec{x})\psi_{jk}(\vec{x}) \diffd \mu(\vec{x})
\end{align*}
so they have many of the same reconstruction properties as orthonormal bases while maintaining additional flexibility (\cite{Intro_frames}). Additionally, the coefficients are asymptotically decorrelated as $j\to \infty$, and due to their local support this holds even if there are large swathes of missing data (\cite{RFSphere}). 
We use the MATLAB package NeedMat to calculate the needlets (\cite{fan}).

\subsubsection{$Z_2(\vec{x})$}\label{sec:residualprocesses}

We consider a suite of common compactly supported options for the small-scale process $Z_2(\vec{x})$. 
The first class of correlation functions we will entertain are the Gaspari-Cohn correlation functions first described in \cite{gasparicohn}. 
This class of correlation functions is generated by a self-convolution of a family of piecewise-linear functions over the domain of interest. 
This family of piecewise-linear functions has the parametrization
\begin{equation}  \label{eq:gcB}
  B_0(d|a,c) = 
  \begin{cases} 
    \frac{ 2(a-1)\abs{d}}{c}+1 &  \abs{d}<c/2 \\
    2a\left(1-\frac{\abs{d}}{c}\right) & c/2 \leq  \abs{d} \leq c \\
    0 & \abs{d} > c 
  \end{cases}
\end{equation}
where $a$ is a shape parameter and $c$ a scale parameter. 
This family exhibits negative correlation for $a=-0.1$, and to distinguish this from the other correlation models we are choosing to fix $a$ to this value (See \cite{gasparicohn} Figure 7). 
The Gaspari-Cohn correlation function is then given by
\begin{equation}
    C(d|a,c) =  B_0(d|a,c) \ast B_0(d|a,c)
\end{equation}
where $\ast$ represents convolution in $\mathbb{R}^n$.

It is important to note that the Gaspari-Cohn correlation function is not positive-definite on the sphere (\cite{gneitingpdsphere}), which means that we must use the chordal distance $d = 2\sin\left(\frac{\alpha}{2}\right)$, where $\alpha = \arccos{\inprod{x_1}{x_2}}$ is the angle between the position vectors of the two points. 
While this can lead to unusual correlation structures over large distances, since the model will be compactly supported and describe local variation we are not concerned with the warping of the surface that this will introduce. 

The second class of covariance models that we consider is a mixture of Wendland covariance functions (\cite{wendlandrbf}). 
The Wendland covariance is a radial basis function with compact support outside a cutoff radius $\theta$ of the form 
\begin{equation}\label{eq:wendlandrbf}
  \phi(d,\theta) = \frac{1}{3}\left(1-\left(\frac{d}{\theta}\right)\right)^6_+
  \left(35\left(\frac{d}{\theta}\right)^2+18\left(\frac{d}{\theta}\right)+3\right).
\end{equation}
Since they are positive definite functions on the sphere (\cite{gneitingpdsphere}), we will use the angular distance measure $d = \arccos{\inprod{x_1}{x_2}}$. 
We propose to model the covariance of the small-scale process using the mixture of Wendland radial basis functions as
\begin{equation}
    C(d) = \sum_{i=1}^n \alpha_i \phi(d,\theta_i)
\end{equation}
for $n = 1,2$. 

Lastly, we investigate the tapered Mat\'ern correlation model discussed in Section \ref{sec:simdata}.
This model is also positive-definite on the sphere, so we will let $d$ be the angular distance.

\subsection{FSBGL Model Fit}

In this subsection we examine the estimated FSBGL model, and illustrate its interpretability with respect to the modeling problem at hand.

\subsubsection{Model Selection}\label{sec:modelselection}

Each model for $Z_2(\vec{x})$ was fit using a range of penalty parameters $\lambda \in [0.002,100]$ and the conditional AIC was calculated for each $\lambda$. As $\lambda$ decreased the change in the conditional AIC became smaller, and we selected the value of $\lambda$ at which the conditional AIC began changing by less than $0.01\%$ per decade. 
The selected conditional AIC and associated trace of the hat matrix is given in Table \ref{tab:cAIC}. 
The tapered Mat\'ern option for $Z_2(\vec{x})$ minimizes the conditional AIC, and additionally had the minimum effective degrees of freedom. 

\begin{table}[]
    \centering
    \begin{tabular}{c|c|c}
         Model& cAIC  &$\tr{\hat{\m{H}}}$ \\
          \hline
         Gaspari-Cohn $a=-0.1$&$1.99 \times 10^5$  & $48.31$\\
          
         1 Wendland & $1.95\times 10^5$&$33.08$ \\
         
        2 Wendlands & $1.93 \times 10^5$& $50.39$\\
        
        Tapered Mat\'ern & $1.86 \times 10^5$ & $22.65$\\
        
    \end{tabular}
    \caption{Conditional AIC and trace of the hat matrix 
    for the models of $Z_2$ in Section \ref{sec:residualprocesses}.}
    \label{tab:cAIC}
\end{table}

The fitted parameters for the four models tested are given in Table \ref{tab:params}. 
The most striking result is the significant variation in the value of the fitted nugget variance $\tau^2$. 
While the tapered Mat\'ern model has a fitted nugget variance of just $0.15$, the fitted variance for other models is at least 100 times larger, and in the case of the Gaspari-Cohn model it is about 650 times larger. 
This variation is likely caused by differences in how well each model can handle the underlying smoothness of the data: a stationary field whose correlation function has $2k$ derivatives at the origin will be $k$ times mean-square differentiable (\cite{steinbook}) and both the Wendland and Gaspari-Cohn correlation functions have derivatives at the origin, implying that the underlying field $Z_2$ has some degree of smoothness. 
However, this smoothness is not supported by the data, and so to compensate for that the nugget variance is driven higher in an attempt to match the observed smoothness. 
Since for the tapered Mat\'ern model we are effectively fitting the number of derivatives of the field the model is able to compensate for this directly, and does, selecting $\nu\approx 0.63$, corresponding to a mean-square continuous but not differentiable $Z_2$. 
This means that the nugget effect is not responsible for matching the underlying smoothness of the field all on its own, and this leads to a much smaller variance being fit. This phenomenon is further investigated in \ref{sec:nuggeffect}

We also see some variation in the fitted widths of the support, most interestingly between the two fitted Wendland models. 
While the single Wendland model ends up with a support width of about $9.25\degrees$, the two range parameters for the mixture Wendland model are about $5\degrees$ and $13\degrees$. 
This suggests that the single Wendland model is trying to average out these dominant scales of variation in $Z_2$, while the mixture Wendland model is able to incorporate them individually. 
This may also explain why the nugget variance is larger in the single Wendland and Gaspari-Cohn models than in the mixture Wendland model, as it may be attempting to fit the smaller scale variations that are not present in the single Wendland and Gaspari-Cohn models, in addition to its role in setting the smoothness of the field. 

\begin{table}[]
    \centering
    \begin{tabular}{c|c|c|c|c|c}
       Model & $\tau^2$ &$\sigma^2$ &Support&Smoothness&Range\\
        \hline
        FSBGL-TM & $0.15$ &$1032$&$27.6\degrees$&$0.63$&$7.5\degrees$\\
        FSBGL-GC & $98.2$ &$858$& $23\degrees$&N/A&N/A \\
        FSBGL-1W & $63.1$& $366$ &$9.25\degrees$ &N/A &N/A \\
        FSBGL-2W & $19.8$& $575$ &$13.2\degrees$ &N/A &$5.1\degrees$
    \end{tabular}
    \caption{Fitted parameters for possible models of $Z_2$. The support parameter is the angular half-width of the support of the covariance function, and the range parameter in the two-Wendland model is the fitted correlation length of small-scale variations.  Models are tapered Mat\'ern (TM), Gaspari-Cohn (GC), single Wendland (1W) and a mixture of two Wendlands (2W).}
    \label{tab:params}
\end{table}

\subsubsection{Structure of $\mQ$}

\begin{figure}
    \centering
    \includegraphics[width=0.45\textwidth]{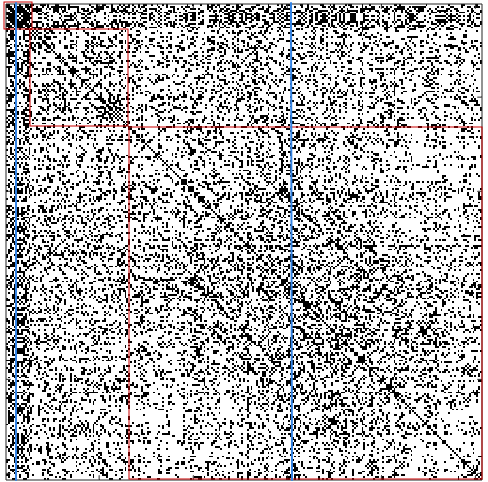}    
    \caption{Sparsity pattern of $\mQ$ fitted with the tapered Mat\'ern model. Zero entries are in white, nonzero in black. Red squares outline the blocks corresponding to same-level dependencies, and the leftmost and rightmost blue lines correspond to the data shown in the left and right columns of Figure \ref{fig:nbhd} respectively.}
    \label{fig:Qsparsity}
\end{figure}

One of the main advantages of a model like the FSBGL for multiresolution modeling is that there is no constraint on the interactions between levels of resolution, unlike models such as LatticeKrig where different levels of resolution are assumed to be independent. 
This allows the different scales of variation to be (potentially) correlated with others. 
As a demonstration, we examine the fitted $\mQ$ using a tapered Mat\'ern model for $Z_2$. 
The sparsity pattern of $\mQ$ is shown in Figure \ref{fig:Qsparsity}, with non-zero entries in black and zero entries in white. 

While the overall graph structure implied by $\mQ$ is not obvious, it is clear that the lowest level of resolution, corresponding to $j=0$, have significant conditional dependencies with higher levels of resolution, while the other two levels of resolution do not show as much interconnection. 
This feature is not possible in models like LatticeKrig, and we illustrate below the superior fit of the the FSBGL.

\begin{figure}
    \centering
    \includegraphics[width=\linewidth]{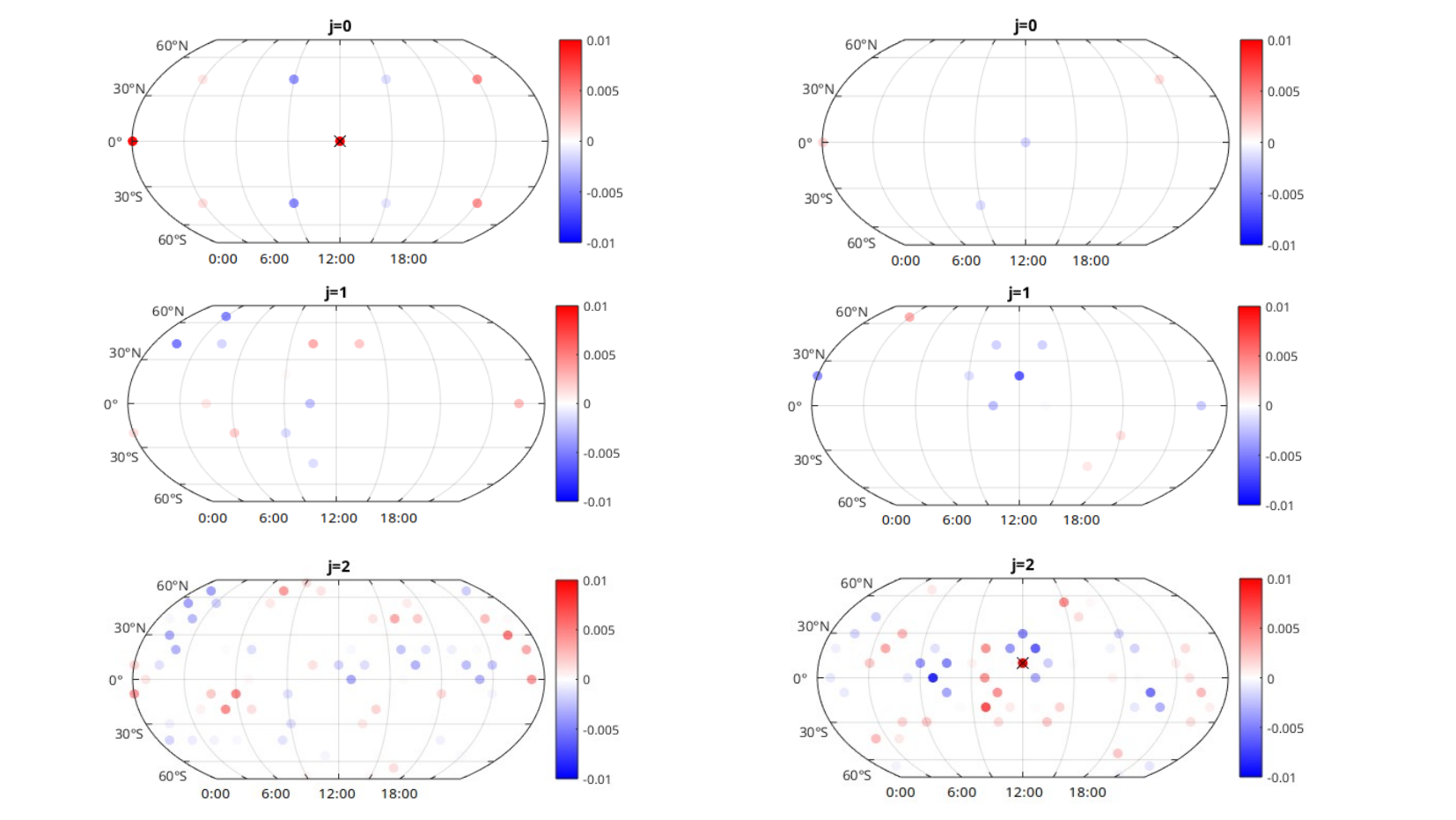}
    \caption{Plots of neighborhood structure for the needlets $\psi_{0,7}$ and $\psi_{2,81}$. The centers of the needlets are marked with a black X.}
    \label{fig:nbhd}
\end{figure}

A closer examination of the neighborhood structure of $\mQ$ is shown in Figure \ref{fig:nbhd}.
In each case, the estimated neighborhood structure is plotted by showing the non-zero entries of $\mQ$ at the nodal points $\xi_{jk}$ of the needlets. 
In the left column we see the neighborhood of the needlet with $j=0,k=7$, the corresponding $\xi_{jk}$ of which is overlaid with a black X located at $0\degrees$ latitude at local noon, and the same in the right column for the needlet with $j=2,k=81$ located at about $10\degrees$ latitude and local noon. 

These figures exhibit patterns that have physical interpretations. 
The left column of Figure \ref{fig:nbhd} shows that $\beta_{0,7}$ has negative partial correlation with the coefficients at $j=0$ located on the night side and positive partial correlation with the needlets with $j=0$ located on the day side. 
This may correspond to the large scale effects of solar heating on the neutral temperature. 
We also note the large number of neighbors in the higher levels of resolution. 
The right column of Figure \ref{fig:nbhd} illustrates that most of the neighbors of $\beta_{2,81}$ are at latitudes less than $30\degrees$, which suggests that there are latitude-dependent dynamics at this scale. 
We also notice that there do not seems to be as many neighbors at the lowest level of resolution, but this does include a positive partial correlation with its dayside neighbors at lower levels of resolution.

\subsubsection{Covariance Structure}

In this section we compare the implied covariance properties of the estimated model to empirical statistics derived from the data. 
While we have a relatively limited number of realizations, which makes the complete determination of the covariance structure directly from the data difficult, we are able to uncover structure that matches broadly with what is seen in the data and what is expected from the known physics of the upper atmosphere.

\begin{figure}
    \centering
    \includegraphics[width=0.75\linewidth]{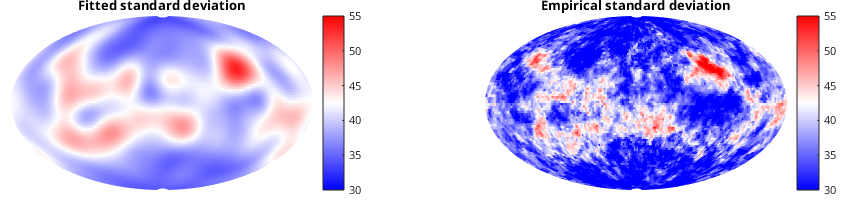}
    \caption{Fitted vs empirical standard deviations for the FSBGL with a Tapered Mat\'ern small-scale model.}
    \label{fig:fitvsemp_stdev}
\end{figure}

We first examine the estimates of the fitted and empirical standard deviations of the data, which are shown in Figure \ref{fig:fitvsemp_stdev}. 
We first notice that the model is able to capture the large-scale structure of the marginal standard deviations fairly well, displaying elevated values in approximately the same areas as the data, albeit without the fine-scale structure. 
However, with such a limited number of samples, much of this very small-scale structure is likely to be noise. 
A noticable potential issue with our methodology is that there are several areas, for example near the poles, where we overestimate the marginal standard deviation. 
This is a consequence of selecting a stationary model for $Z_2$, in that we cannot capture spatial variation in the marginal standard deviation that is not already captured by the low-rank portion. 
A situation such as the one observed here could be remedied by allowing the marginal variance of $Z_2$ to be some function of latitude, for example $\sigma^2(\phi)=\alpha_1+\alpha_2\cos\phi$, with $\alpha_1$ and $\alpha_2$ fit along with the other parameters, and more general situations handled with different functional forms, but our comparisons below suggest this more parsimonious formulation still describes the data better than competing models.

\begin{figure}
    \centering
    \includegraphics[width=0.75\linewidth]{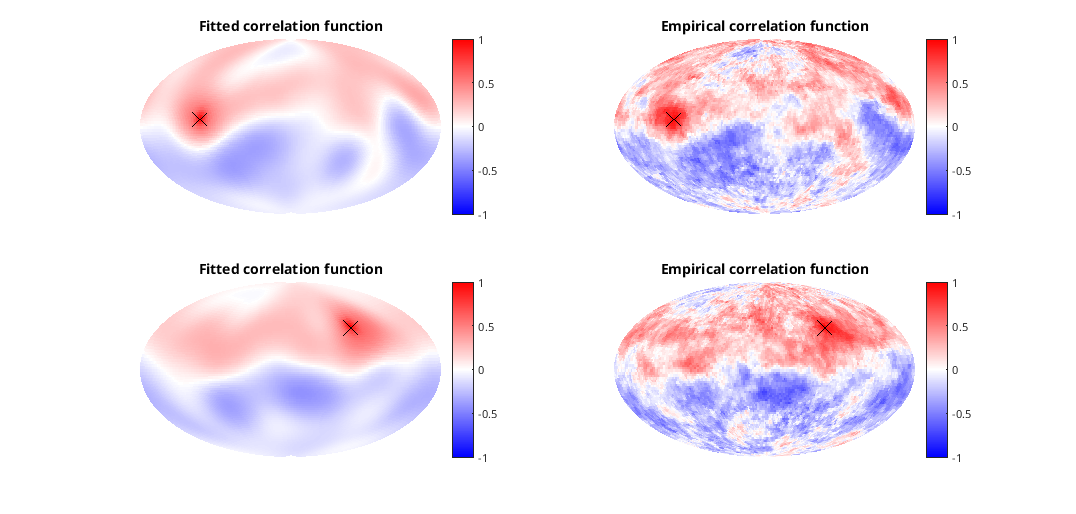}
    \caption{Fitted vs empirical correlation functions for the FSBGL with a tapered Mat\'ern small-scale model. The black X marks the center of the correlation functions.}
    \label{fig:fitvsemp_corr}
\end{figure}

We also consider fitted and empirical correlation functions. 
We developed this model under the assumption that the local correlation structure would be approximately stationary and that most nonstationarity would be in the large-scale structures, and we see this reflected in the empirical correlation functions (Figure \ref{fig:fitvsemp_corr}).  
While there is local variability in the empirical correlation function at long ranges, we expect that most of this is noise. 
We can see that the local correlation structure is well-represented by the model, and that we are also able to capture the broader correlation structure on large scales. 
An important feature of the data that we see reproduced in the model is the broad anticorrelation between hemispheres; this is due to seasonality and that we are able to detect it without the statistical model having prior knowledge of its existence is an encouraging result. 

\subsection{Comparison with Extant Approaches}\label{sec:extant_comp}

We compare the performance of the FSBGL to existing models within a similar framework. 
The models chosen for comparison are LatticeKrig, the basis graphical lasso (BGL), and the full-scale approximation. 
The next subsections detail the specifications of these competing alternatives.

\subsubsection{LatticeKrig}

The LatticeKrig model introduced in \cite{nychka} is a multiresolution model in which the field is described via Wendland radial basis functions and the coefficients are modeled as a Gaussian Markov random field with node locations at the centers of the radial basis functions. 
By describing the coefficients with a spatial autoregression, $\mQ$ is guaranteed to be a sparse matrix, and the compact support of the basis functions introduces sparsity in the basis matrix $\m{\Psi}$ as well. 
This sparsity can be leveraged to allow the model to be evaluated extremely quickly, making likelihood computations fast and minimizing the computational load of model fitting. 
In addition, the structure allows the LatticeKrig model to adapt to a wide variety of spatial covariance functions (\cite{nychka}). 
In a spherical geometry the LatticeKrig model has a similar number of basis functions to the needlet frame at each level of resolution, so we assume three levels of resolution as we did for the FSBGL. 

LatticeKrig is a stationary model and due to its parametrization cannot capture the full variability present in the data. 
This means that it performs somewhat more poorly in this context than the methods we have discussed, which are able to infer some level of nonstationarity through the data to inform the structure of $\mQ$. 
The negative log-likelihood of LatticeKrig  on the test set to is approximately $1.36 \times 10^5$, far higher than any of the other models (see Table \ref{tab:outside}). 
As will be discussed in more detail later, it is also the poorest predictor of subgrid-scale data as it is not able to leverage any knowledge of nonstationarity.

Due to the structure of the R package, LatticeKrig must be run with Wendland basis functions. The work in \cite{bgl} showed that the BGL significantly outperformed the LatticeKrig model using Wendland functions, and we will see that our model outperforms the BGL when both use needlets. For completeness, we have included a comparison of the FSBGL using both a needlet and Wendland basis in \ref{sec:basis_comp}, along with the differences in the fitted parameters, likelihoods, and predictive performance. We show that the FSBGL performs nearly identically if either basis is used, providing further evidence that the differentiating factor is the inclusion of the small scale structure rather than the basis. 

\subsubsection{Basis Graphical Lasso}

The principal difference between our construction and that of the basis graphical lasso lies in the process $Z_2$. 
While the basis graphical lasso assumes that $Z_2=0$, our extension has generalized this to a stationary, compactly supported covariance. 
This allows much greater flexibility in the modeling of spatial datasets with multiscale structure, since, as noted in \cite{steinlowrank}, low-rank models like the basis graphical lasso have trouble recreating small-scale variability. 
Additionally, when applied to the test dataset, while the BGL is clearly superior to LatticeKrig, it is outclassed by the other models by a similar amount. 

\subsubsection{Full-Scale Approximation}

This model is motivated by the full-scale approximations (FSA) of \cite{steinfullscale} and \cite{sangfullscale}, so we wish to compare the results of applying both methods. 
Due to the limitations of the model proposed by \cite{steinfullscale}, where the entire covariance matrix $\mQ^{-1}$ is estimated, severely limiting the number of basis functions, we will be following the specification laid out in \cite{sangfullscale}. 
There, the low-rank model consists of EOFs calculated from the same 32 realizations as the other algorithms presented, and we selected the EOFs that explained $70$ percent of the variance seen in the data. 
This value was chosen by examination of the fitted low-rank portion of the FSBGL, where the low-rank portion explained less than $70\%$ of the variance at $90\%$ of spatial locations. 
Since the EOFs are generated from the data, we expect to see that they perform the best on the in-sample data. 
Additionally, calculating the EOFs is fast, and their orthonormality simplifies the computations of the parameters of $C_2$. 

\begin{table}[]
    \centering
    \begin{tabular}{c|c}
       Model & Negative log-likelihood (test set) \\
        \hline
        LatticeKrig & $1.36 \times 10^5$ \\
        BGL & $1.21 \times 10^5$  \\
        Full-Scale & $ 1.03\times 10^5$   \\
        FSBGL-GC & $1.02 \times 10^5$   \\
        FSBGL-1W &$1.01 \times 10^5$  \\
        FSBGL-2W & $0.99 \times 10^5$   \\
        FSBGL-TM & $\bf{0.96\times 10^5}$
    \end{tabular}
    \caption{Negative log-likelihoods 
    on out of sample data.}
    \label{tab:outside}
\end{table}

One of the drawbacks of EOF models is that they often overfit training data. 
Table \ref{tab:outside} shows the FSA is  worse than the FSBGL in terms of fitting data outside the training set, suggesting that the fitted FSBGL generalizes better. 

\subsubsection{Timing Study}

The combination of the FSA and BGL does not induce a limiting computational burden beyond either of the component models. 
Table \ref{tab:runtime} compiles representative training times per iteration for several of the models examined in this paper. 
Calculations were done on the National Center for Atmospheric Research's Derecho supercomputer using 32 CPUs and 55 GB of memory. 
All code was written in MATLAB, and for the purposes of the timing study the simulated annealing algorithm was initialized to the center of the parameter search grid and $\lambda$ was set to 1. 
The simulated annealing algorithm is considered to have converged when 20 consecutive steps fail to change the value of the log-likelihood by more than $0.1\%$. 
Due to the dependence of simulated annealing on the random steps taken, we have reported the time in seconds per iteration to achieve convergence. 
This typically occurred in 200-300 steps, and the calculation of $\mQ$ typically converged in 15-20 steps. 
For the BGL, the values of $\hat{p}$ were calculated in about 30 iterations. 

\begin{table}[]
    \centering
    \begin{tabular}{c|c|c}
       Model & Calculating $\hat{p}$ (s/iteration)& Calculating $\mQ$ (s/iteration) \\
        \hline
        BGL & 0.1 & 26 \\
        Full-Scale & 6.5& 0.62  \\
        FSBGL-GC &5.1&32 \\
        FSBGL-1W & 2.8 & 24\\
        FSBGL-2W &7.4 & 39\\
        FSBGL-TM & 6.2&21
    \end{tabular}
    \caption{Training time per iteration for each model, seconds. Calculating $\hat{p}$ typically takes 200-300 iterations for the FSBGL and calculating $\mQ$ takes 15-20. The BGL typically achieves convergence in about 30 iterations of each algorithm, and since the full-scale approximation uses SVD to calculate $\mQ$ it takes one iteration.}
    \label{tab:runtime}
\end{table}

Since the BGL is able to be run using the \texttt{fmincon} function in MATLAB, which utilizes BFGS in a trust-region reflective algorithm, rather than simulated annealing for the other models, the calculation of $\hat{p}$ is much faster than for any of the other models. 
In spite of this, $\hat{p}$ only needs to be calculated once since the optimization problem in \eqref{eq:invertD} is independent of the value of $\lambda$. 
This means that determining $\mQ$ for different values of $\lambda$ only requires minimizing \eqref{eq:prop1} again via the algorithm in \eqref{eq:dcupdate}, and since typical uses will involve calculating $\mQ$ for a range of $\lambda$ values this additional time is not of great concern. 
Similarly, since the calculation of $\mQ$ for the full-scale approximation involves only an SVD of the data matrix, that calculation is by far the fastest.

We also want to note that the time to determine $\mQ$ is very sensitive to the value of $\lambda$. 
Results here are shown for $\lambda=1$, however smaller values of $\lambda$ will result in less sparsity in $\mQ$, making the calculation take longer. 
Similarly, larger values of $\lambda$ result in more sparsity, making the calculation faster until the point where the fitted $\mQ$ is diagonal. 

\subsubsection{Prediction}

While the model was fit on data with a $2\degrees$ resolution in latitude and longitude, the model can be run at higher resolutions, for example we have limited access to $0.25\degrees$ resolution data. 
We consider this subgrid-scale data to test the predictive performance of the model by calculating the continuous rank probability score (CRPS) and the root mean square error (RMSE). 
Reported CRPS and RMSE values are averaged over individual predictions; CRPS has a closed form for Gaussian predictive distribution (\cite{scoring}), 
\begin{equation}
  CRPS(F) = -\sigma \left[ \frac{1}{\sqrt{\pi}} - 2\phi\left(\frac{y-\mu}{\sigma}\right) - \left(\frac{y-\mu}{\sigma}\right)\left(2\Phi\left(\frac{y-\mu}{\sigma}\right)-1\right) \right]
\end{equation}
where $\phi \ \text{and}\ \Phi$ are the standard normal PDF and CDF respectively, $F$ is the normal predictive distribution with mean $\mu$ and variance $\sigma^2$ and $y$ is the realizing observation. 

The subgrid-scale data was predicted at approximately 5000 randomly selected spatial locations, with the predictions done independently at these locations for each of the 41 fields. 
This leads to approximately 200,000 total predictions. 
Validation scores for LatticeKrig, the BGL and FSBGL models are shown in Table \ref{tab:predictions}. 

\begin{table}[]
    \centering
    \begin{tabular}{c|c|c|c}
         Model& Mean CRPS & Median CRPS & RMSE \\
         \hline
         LatticeKrig & 16.71 &11.57&28.89\\
         BGL & 13.47& 9.57&23.98 \\
         FSBGL-GC&6.52&3.96&10.70\\
         FSBGL-1W&5.72&3.57&10.61\\
         FSBGL-2W&5.68&3.31&10.74\\
         FSBGL-TM&5.58&3.15&10.51
    \end{tabular}
    \caption{Mean and median CRPS, as well as RMSE, for subgrid-scale predictions derived using the fitted models.}
    \label{tab:predictions}
\end{table}

From these results, it is clear that predictions are much more accurate using the FSBGL with the mean and median CRPS, as well as the RMSE, being less than half that of the BGL and LatticeKrig. 
In fact, the lowest CRPS obtained by LatticeKrig at \textit{any} predicted location is 4.64, which is more than the median CRPS for any of the models fitted with the FSBGL. 
This suggests that the FSBGL is also suitable to be used to fit predictive distributions to data exhibiting multiscale behavior, for example one potential route of future investigation could be whether this technique can be allow for running climate models at lower spatial resolution from which FSBGL can statistically downscale to a fine resolution. 
This should be especially contrasted with the full-scale approximations of \cite{sangfullscale}, for example, as the reliance on EOFs for the low-rank portion of the model makes subgrid-scale prediction impossible. 

\section{Discussion}\label{sec:fsbgl_discussion}

The FSBGL provides a flexible and computationally tractable way to model large spatial datasets. By combining a nonstationary low-rank model with a sparse precision matrix via a lasso regularization and a stationary model with compactly supported covariance function the method is able to accurately represent the covariance structure of complex spatial datasets. Due to its flexibility and general structure, it is able to be adapted to a wide variety of applications. By not specifying the basis functions, as done in LatticeKrig or in the full-scale approximations of \cite{sangfullscale}, the modeler is given the ability to use their domain expertise to choose the low-rank representation they prefer. This could be wavelets, as done here, radial basis functions, or an orthonormal basis, among other options. While \cite{steinfullscale} also developed their model in this way, it is computationally infeasible to employ more than a few basis functions in the low-rank portion due to the need to estimate the Cholesky decomposition of the coefficient covariance matrix. However our inclusion of the lasso penalty allows much more expressive low rank representations to be applied due to the enforced sparsity of the estimated $\m{Q}$. The modeler additionally has wide latitude in selecting their residual model, with the only requirement being that it is compactly supported. 

One area where the method can be adapted is to allow for non-stationarity in $Z_2$. For example, it can be seen in Figure \ref{fig:fitvsemp_stdev} that the empirical standard deviation of the dataset is highly spatially variable. The easiest way to adapt the model to handle nonstationary residuals would be to profile the variance by having the residual variance be some function of a set of spatial predictors. Such an approach was used in \cite{axingneed} to model nonstationarity in the ionospheric electrostatic potential and shown to work well. Another option would be to choose a model that is naturally nonstationary, such as the generalized Gaspari-Cohn model \cite{gilpin}. Due to the highly spatially variable influence of atmospheric waves, which provide a large portion of the energy input into the lower thermosphere, this seems to be a natural avenue for further study in this application \cite{liu.gravity.waves}. 

In addition to their spatial variability, atmospheric waves also impart significant temporal variability on neutral temperature fields, especially in the lower thermosphere.  It is helpful to incorporate the dependence of temporal scales on their spatial scales. For this reason, future work could involve extending these ideas into a time-dependent scenario. There are several ways this could be done. Autoregressive structure in the basis coefficients could be endowed, or sets of spatiotemporal functions such as radial 3D needlets could be investigated (\cite{threedneedlets}). Alternatively, it may be possible to leverage a software package such as EiGLasso (\cite{eiglasso}) or TeraLasso (\cite{teralasso}) to determine a Kronecker sum structure using an algorithm similar to the one discussed here. 

Another avenue for further investigation involves the use of the model as a statistical downscaler. The subgrid-scale prediction results suggest that the model is better at predicting localized variation, both in terms of CRPS and RMS error, than other spatial Gaussian process models. Since geophysical model runs such as WACCM-X at 0.25 degree resolution are very computationally intense, being able to run the model at low resolution and perform conditional simulation to estimate the subgrid scale behavior and associated uncertainties may be useful. While the results of the previous section are encouraging, whether this model is able to do this estimation operationally and with sufficient accuracy to justify its use in this way remains to be seen.



\section{Conclusion}

This paper proposes the synthesis of two state-of-the-art methods for spatial data, the basis graphical lasso and full-scale approximation, for modeling of nonstationary processes where estimation and inference is challenged by high spatial resolution.
The field is modeled as consisting of a low-rank component with coefficient precision matrix determined by an $\ell_1$ penalized DC algorithm and a small-scale component, which, unlike the basis graphical lasso, is described by a parametric covariance function instead of a nugget effect alone. 
We showed that the method is able to recover known structure in the precision matrix of the coefficients and able to estimate the parameters of the residual field even with only a few tens of replicates of the field available.
We also examined the performance of this algorithm as a model of neutral temperature fields for a variety of commonly used compactly supported covariance functions and compared them to existing models for large spatial datasets in terms of likelihood and prediction, as well as an examination of samples from the fitted distributions. 
We found that this method demonstrated significantly better performance on held out data both in terms of subgrid scale prediction accuracy and likelihood, and was able to capture salient features of the dataset. 
We also showed that draws from the fitted distribution were sufficiently similar to the data that the method can be used to model complex geospatial datasets. 
This suggests that the method may be able to augment ensemble whole-atmosphere simulations to help engineers account for the effects of spatial structure in the thermosphere when modeling vehicles in reentry.

\if0\blind
{
\section{Supplementary Materials}
The code used in this project is available at \texttt{https://github.com/mfleduc/FullScaleBGL} along with some sample datasets.

\section{Funding}

This research was supported by NASA Award number 80NSSC22K0175, NSF AGS-1848544, and NSF DMS-2310487.
}\fi

\newpage
\appendix
\section{Proofs and Derivations}
\subsection{Proof of Proposition \ref{thm:prop1}}

The likelihood given by \eqref{eq:bgllike} is
\begin{equation}
     L(\mQ | \mD,\m{Y}) =  \log \det\left(\m{\Psi} \mQ^{-1} \m{\Psi}^T+\mD \right) + \\  \text{tr}\left(\m{S}\left(\m{\Psi} \mQ^{-1} \m{\Psi}^T+\mD \right)^{-1}\right) + ||\m{\Lambda} \circ \mQ||_1
\end{equation}
Following the Matrix Determinant Lemma, we have that 

\begin{equation}
            \log\det(\mD+\m{\Psi} \mQ^{-1}\m{\Psi}^T) = \log\det(\mQ+\m{\Psi}^T \mD^{-1}\m{\Psi}) + \log\det(\mD) - \log\det(\mQ)
\end{equation}
and by the Woodbury Formula, 
\begin{equation}
    (\mD+\m{\Psi} \mQ^{-1}\m{\Psi}^T)^{-1} = \mD^{-1}-\mD^{-1}\m{\Psi}(\mQ+\m{\Psi}^TD^{-1}\m{\Psi})^{-1}\m{\Psi}^T\mD^{-1}
\end{equation}
Applying the above and the properties of trace, we find that 
\begin{equation}
    \begin{split}
        L(\mQ | \mD, \m{Y}) = \log\det(\mQ+\m{\Psi}^T \mD^{-1}\m{\Psi}) + \log\det(\mD) - \log\det(\mQ) \\ +\tr{\m{S D}^{-1}} -\tr{\m{\Psi}^T \mD^{-1}\m{S D}^{-1} \m{\Psi}\left(\mQ+\m{\Psi}^T \mD^{-1} \m{\Psi}\right)^{-1}} + ||\m{\Lambda} \circ \mQ||_1
    \end{split}
\end{equation}

This is the likelihood used to fit the small-scale model when $\m{\Lambda}=\m{0}$. Removing the terms that do not depend on $\mQ$ gives the expression of the likelihood in Proposition \ref{thm:prop1}. 

\subsection{Derivation of Equation (\ref{eq:dcupdate})}

Starting from the expression of the likelihood given in \eqref{eq:prop1}, we note that $\norm{\m{\Lambda} \circ \mQ}_1$ is a convex function of $\mQ$, and by \cite{convexoptimization} so is $-\log\det(\mQ)$. By the same logic, $\log\det(\mQ + \m{\Psi}^T\mD^{-1}\m{\Psi})$ is concave in $\mQ$. Also by \cite{convexoptimization}, for positive definite $\m{\Sigma}$, $\tr{\m{S\Sigma}^{-1}}$ is convex. So, the problem can be separated into a convex portion given by $f(\mQ) = -\log\det(\mQ) + \norm{\m{\Lambda} \circ \mQ}_{1}$ and a concave portion given by $g(\mQ) = \log\det(\mQ + \m{\Psi}^T\mD^{-1}\m{\Psi})-\tr{\m{\Psi}^T \mD^{-1}\m{S D}^{-1} \m{\Psi}\left(\mQ+\m{\Psi}^T \mD^{-1} \m{\Psi}\right)^{-1}}$. This gives the likelihood function the form of a difference of two convex functions, more explicitly: 

\begin{equation}\label{eq:concavevex}
\begin{split}
     L(\mQ | \mD, \m{Y}) = &\underbrace{\log\det(\mQ+\m{\Psi}^T \mD^{-1}\m{\Psi}) -\tr{\m{\Psi}^T \mD^{-1}\m{S D}^{-1} \m{\Psi}\left(\mQ+\m{\Psi}^T \mD^{-1} \m{\Psi}\right)^{-1}}}_{concave = g(\mQ)}
\\ &\underbrace{- \log\det(\mQ)  + ||\m{\Lambda} \circ \mQ||_1}_{convex = f(\mQ)}
     \end{split}
\end{equation}

Now, as done in \cite{bgl}, we can apply the Difference of Convex algorithm (\cite{DCA}). To do this we note that since $g(\mQ)$ is concave, it is bounded above by its linearization $\Tilde{g}(\mQ|\mQ_j)$ given by

\begin{equation}\label{eq:gQmajorant}
   \Tilde{g}(\mQ|\mQ_j) = g(\mQ_j) + \inprod{\nabla g(\mQ_j)}{\mQ-\mQ_j} = g(\mQ_j) + \tr{\nabla g(\mQ_j)(\mQ-\mQ_j) }
\end{equation}
with equality iff $\mQ=\mQ_j$. Since we define $\mQ_{j+1} =\underset{\mQ>0}{\text{argmin}}\left[ f(\mQ)+\Tilde{g}(\mQ|\mQ_j) \right ]$, we know that 
\begin{equation}
     L(\mQ_{j+1} | \mD, \m{Y}) \le f(\mQ_{j+1})+ \Tilde{g}(\mQ_{j+1}|\mQ_j) \le f(\mQ_{j})+ \Tilde{g}(\mQ_{j}|\mQ_j) =  L(\mQ_{j} | \mD, \m{Y}) 
\end{equation}
and so iteratively minimizing $\Tilde{L}(\mQ|\mQ_j,\mD,\m{Y}) = f(\mQ)+\Tilde{g}(\mQ|\mQ_j)$ will minimize $L(\mQ|\mD,\m{Y})$ as well. Removing the terms in $\Tilde{g}(\mQ|\mQ_j)$ that do not depend upon $\mQ$, we see that 

\begin{equation}
    \hat{\mQ}_{j+1} = \underset{\mQ>0}{\text{argmin}} [ \Tilde{L}(\mQ|\hat{\mQ}_j,\mD,\m{Y}) ]= \underset{\mQ>0}{\text{argmin}}\left[- \log\det(\mQ) +\tr{ \nabla g(\hat{\mQ}_j) \mQ} + ||\m{\Lambda} \circ \mQ||_1  \right]
\end{equation}
as desired. 

Now it remains to calculate $\nabla g(\mQ)$. First we can note that via the chain rule

\begin{equation}
    \pderiv{}{\mQ_{ij}}\log\det(\mQ+\m{\Psi}^T \mD^{-1} \m{\Psi}) = \frac{1}{\det(\mQ+\m{\Psi}^T \mD^{-1} \m{\Psi})}\pderiv{}{\mQ_{ij}}\det(\mQ+\m{\Psi}^T \mD^{-1} \m{\Psi})
\end{equation}
so by Jacobi's Formula we get that 
\begin{equation}
    \label{eq:ldpartgradg}
    \pderiv{}{\mQ_{ij}}\log\det(\mQ+\m{\Psi}^T \mD^{-1} \m{\Psi}) = (\mQ+\m{\Psi}^T \mD^{-1} \m{\Psi})^{-1}.
\end{equation}
Now we must calculate $\pderiv{}{\mQ_{ij}}\tr{\m{\Psi}^T \mD^{-1}\m{S D}^{-1} \m{\Psi}\left(\mQ+\m{\Psi}^T \mD^{-1} \m{\Psi}\right)^{-1}}$. To do this, we let $\m{M}=\left(\mQ+\m{\Psi}^T \mD^{-1} \m{\Psi}\right)^{-1}$ and $\m{B} = \m{\Psi}^T \mD^{-1}\m{S D}^{-1} \m{\Psi}$. Since $\pderiv{\m{I}}{\m{Q}_{ij}} = \m{0} = \pderiv{}{\m{Q}_{ij}}\m{MM}^{-1}$, applying the product rule we see that 
\begin{equation}
    \pderiv{\m{M}}{\mQ_{ij}} = -\m{M}\left(\pderiv{\m{M}^{-1}}{\mQ_{ij}}\right)\m{M}
\end{equation}
Since $\m{M}^{-1} = \mQ+\m{\Psi}^T \mD^{-1} \m{\Psi}$ we can see that $\pderiv{\m{M}^{-1}}{\mQ_{ij}} = \m{\chi}_{ij}$, the matrix of all zeros except a $1$ in the $i,j$ entry. Since the trace is a linear operator it commutes with the derivative, and direct calculation shows that 
\begin{equation}
    \pderiv{}{\mQ_{ij}}\tr{\m{BM}} = -\tr{\m{BM\chi}_{ij}\m{M}} = -(\m{MBM})_{ij}
\end{equation}
and thus 
\begin{equation}
    \label{eq:tracepartgradg}
    \begin{split}
    \pderiv{}{\mQ_{ij}}\tr{\m{\Psi}^T \mD^{-1}\m{S D}^{-1} \m{\Psi}\left(\m{Q}+\m{\Psi}^T \mD^{-1} \m{\Psi}\right)^{-1}}\\ =-(\mQ+\m{\Psi}^T\mD^{-1}\m{\Psi})^{-1}(\m{\Psi}^T\mD^{-1}\m{SD}^{-1}\m{\Psi})(
    \mQ+\m{\Psi}^T\mD^{-1}\m{\Psi})^{-1}
    \end{split}
\end{equation}
Combining this with the result in \eqref{eq:ldpartgradg}, we see that 
\begin{equation}
    \nabla g(\mQ) =(\mQ+\m{\Psi}^T \mD^{-1}\m{\Psi})^{-1}(\mQ+\m{\Psi}^T\mD^{-1}\m{\Psi}+  \m{\Psi}^T\mD^{-1}\m{SD}^{-1}\m{\Psi})(\mQ+\m{\Psi}^T\mD^{-1}\m{\Psi})^{-1}
\end{equation}
as desired.

\newpage
\section{Simulation Studies}\label{sec:simdata_appendix}

\subsection{Block-Diagonal model for $\mQ$}\label{sec:simdata_appendix_bd}

\begin{figure}[h]
    \centering
    \includegraphics[width=0.85\linewidth]{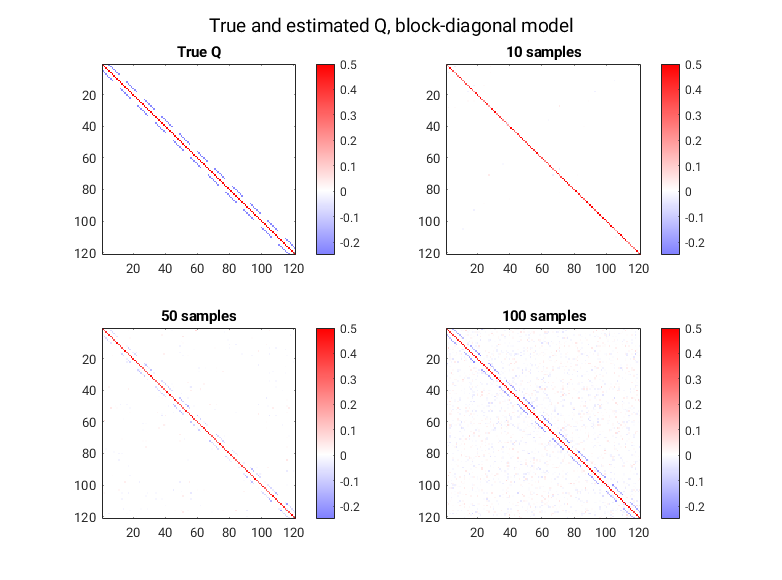}
    \caption{True and estimated $\mQ$ for the block-diagonal model. The true $\mQ$ is in the top left, and the remaining figures show the recovered $\mQ$ with 10, 50, and 100 samples of the field.}
    \label{fig:true_and_est_Q_bdiagmodel}
\end{figure}

\begin{table}[h!]
    \centering
    \begin{tabular}{|c|c|c|c|c|c|}
    \hline
        Parameter&Truth & N=10 &N=30  &N=50& N=100  \\
          \hline
    $\mQ$ & block-diag & 0.610 &0.444 &0.436 & 0.277\\      
    \hline
     $\mQ^{-1}$ & block-diag & 0.654 &0.515 &0.421 & 0.435\\ \hline
    Missed non-zero (\%) &N/A& 51.6&13.8&8.73&0 \\
    \hline
    Missed zero (\%) &N/A& 0.25&4.8&1.54&19.2 \\
    \hline
    $\sigma^2$&1&0.041 & 0.041&0.059 &0.041\\
    \hline
    $\log_{10}(\tau^2)$&-2& 0.224&0.184& 0.232&0.236\\
    \hline
    $\theta$&0.3&0.027& 0.019& 0.043& 0.035\\
    \hline
    $r$& 0.15&0.027& 0.043& 0.036& 0.044\\
    \hline
    $\nu$& 0.5&0.043& 0.073& 0.062& 0.060\\
    \hline
    $L(\hat{Q},\hat{p})/L(Q,p)$ &1&1.1&1.1&1.04&1.1\\
    \hline
    \end{tabular}
    \caption{Median error in recovery of residual parameters for varying number of independent replicates of the field using a block-diagonal model for $\mQ$, along with true values. All calculations are the median over 50 trials for a given number of independent replicates of the field. The error in the recovery of $\mQ$ is given by $\|\hat{Q}-Q\|/\|Q\|$ with $\|\cdot\|$ the Frobenius norm, and the error for the other parameters is the absolute error.}
    \label{tab:simresults_bdiag_appendix}
\end{table}

\begin{table}[h!]
    \centering
    \begin{tabular}{|c|c|c|c|c|c|c|}
    \hline
         $\lambda$&Parameter&Truth& N=10&N=30&N=50&N=100  \\
         \hline
        \multirow{4}{4em}{$0.1$}&$\mQ$&block-diag&9.35&3.32&1.76& 0.733 \\
        &$\mQ^{-1}$&block-diag&2.88 &1.54 &1.16 &  0.795 \\
          &Missed non-zero (\%) &N/A& \textbf{21.1}& \textbf{2.91}&\textbf{0} & \textbf{0} \\
    
    &Missed zero (\%) &N/A& 20.5& 44.4&52.5 & 54.2 \\
        \hline
           
        \multirow{4}{4em}{$0.31623$}&$\mQ$&block-diag&2.67&0.821&0.468&\textbf{0.277}  \\
        &$\mQ^{-1}$&block-diag& 2.51&1.19 & 0.80&   0.435\\
          &Missed non-zero (\%) &N/A& 21.8& 4.36&\textbf{0} & \textbf{0} \\
    
    &Missed zero (\%) &N/A& 18.2& 28.2& 26.9&19.2  \\
         \hline
        \multirow{4}{4em}{$1$}&$\mQ$&block-diag&0.679&\textbf{0.444}&\textbf{0.436}&  0.4323\\
        &$\mQ^{-1}$&block-diag&1.51 &\textbf{0.515}& \textbf{0.421}&   \textbf{0.397}\\
          &Missed non-zero (\%) &N/A& 26.9&13.8 &8.73 &4.36  \\
    
    &Missed zero (\%) &N/A& 11.1& 4.82& 1.54&  0.100\\
       
          \hline
        \multirow{4}{4em}{$3.1623$}&$\mQ$&block-diag&\textbf{0.610}&0.544&0.539& 0.533 \\
        &$\mQ^{-1}$&block-diag& 0.654&0.571 & 0.548&   0.538\\
          &Missed non-zero (\%) &N/A&   51.6& 56.0&56.0 & 56.0\\
    
    &Missed zero (\%) &N/A&   0.25& \textbf{0}& \textbf{0} &\textbf{0}\\
        
         \hline
         
        \multirow{4}{4em}{$10$}&$\mQ$&block-diag&0.614&0.542&0.539&0.533  \\
        &$\mQ^{-1}$&block-diag&\textbf{0.651}&0.573 & 0.548&   0.538\\
          &Missed non-zero (\%) &N/A& 56.0& 55.3& 56.0& 56.0 \\
    
    &Missed zero (\%) &N/A& \textbf{0}& \textbf{0}&\textbf{0} & \textbf{0} \\
        \hline
    \end{tabular}
    \caption{Median statistics for error in recovery of $\mQ$ and the sparsity pattern of $\mQ$ for varying $\lambda$ under a block-diagonal model. Median over 50 independent trials. Note that the parameters of the process $Z_2$ are not included here, since they are inferred before $\mQ$ and independent of $\lambda$. Best estimates for each value of $N$ are bolded.}
    \label{tab:med_errs_lambda_blockdiag}
\end{table}

\newpage
\subsection{Hub and Spoke model for $\mQ$}

\begin{figure}[h]
    \centering
    \includegraphics[width=0.85\linewidth]{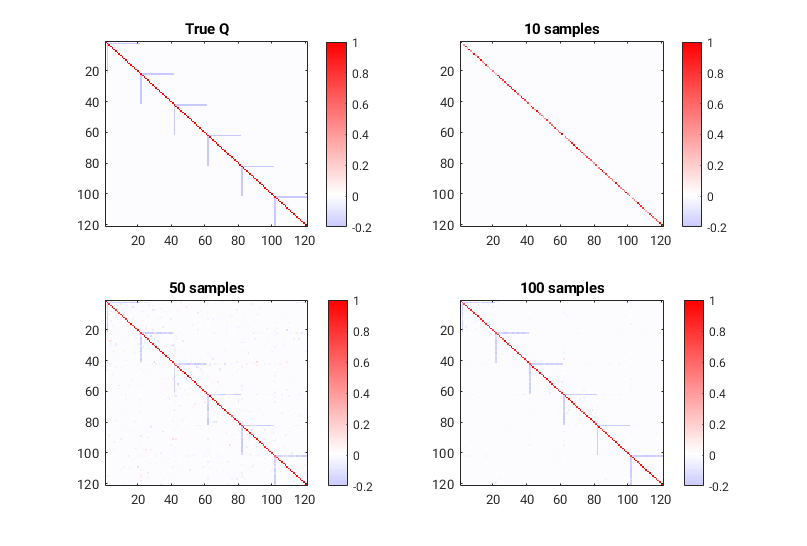}
    \caption{True and estimated $\mQ$ for the hub and spoke model. The true $\mQ$ is in the top left, and the remaining figures show the recovered $\mQ$ with 10, 50, and 100 samples of the field.}
    \label{fig:true_and_est_Q_hubandspoke}
\end{figure}

\begin{table}[h!]
    \centering
    \begin{tabular}{|c|c|c|c|c|c|}
    \hline
        Parameter&Truth & N=10 &N=30  &N=50& N=100  \\
          \hline
    $\mQ$ & hub and spoke & 0.487 &0.336 &0.240 & 0.181\\      
          \hline
           $\mQ^{-1}$& hub and spoke&0.901 &0.552 &0.470 &0.310   \\ \hline
    Missed non-zero (\%) &NA& 12.0&1.15&0&0 \\
    \hline
    Missed zero (\%) &NA& 3.40&1.22&8.27& 4.1\\
    \hline
    $\sigma^2$&1&0.059 & 0.059&0.041 &0.059\\
    \hline
    $\log_{10}(\tau^2)$&-2& 0.342&0.222& 0.288&0.270\\
    \hline
    $\theta$&0.3&0.035& 0.039& 0.021& 0.027\\
    \hline
    $r$& 0.15&0.051& 0.052& 0.045& 0.037\\
    \hline
    $\nu$& 0.5&0.082& 0.084& 0.071& 0.059\\
    \hline
    $L(\hat{\mQ},\hat{p})/L(\mQ,p)$ &1&1.20&1.13&1.12&1.08\\
    \hline
    \end{tabular}
    \caption{Median error in recovery of residual parameters for varying number of independent replicates of the field using a hub-and-spoke model for $\mQ$, along with true values. All calculations are the median over 50 trials for a given number of independent replicates of the field. The error in the recovery of $\mQ$ is given by $\|\hat{\mQ}-\mQ\|/\|\mQ\|$ with $\|\cdot\|$ the Frobenius norm, and the error for the other parameters is the absolute error.}
    \label{tab:simresults_has_appendix}
\end{table}
\begin{table}[h!]
    \centering
    \begin{tabular}{|c|c|c|c|c|c|c|}
    \hline
         $\lambda$&Parameter&Truth& N=10&N=30&N=50&N=100  \\
         \hline
        \multirow{4}{4em}{$0.1$}&$\mQ$&hub and spoke&4.90&1.58&0.856& 0.397 \\
        &$\mQ^{-1}$&block-diag& 1.28&0.739 & 0.563&   0.370\\
          &Missed non-zero (\%) &N/A& 12.6& 2.29&\textbf{0} & \textbf{0} \\
    
    &Missed zero (\%) &N/A& 18.9& 34.1&35.7 & 30.2 \\
        \hline
           
        \multirow{4}{4em}{$0.31623$}&$\mQ$&hub and spoke&1.24&0.367&\textbf{0.240}&\textbf{0.181}  \\
        &$\mQ^{-1}$&block-diag& 1.15& 0.630&\textbf{ 0.470}&   \textbf{0.310}\\
          &Missed non-zero (\%) &N/A& \textbf{12.0}& \textbf{0.57}&\textbf{0} & \textbf{0} \\
    
    &Missed zero (\%) &N/A& 13.5& 12.1& 8.27&4.10  \\
         \hline
        \multirow{4}{4em}{$1$}&$\mQ$&hub and spoke&\textbf{0.487}&\textbf{0.336}&0.331&  0.330\\
        &$\mQ^{-1}$&block-diag& 0.901&\textbf{ 0.552}& 0.474&   0.401\\
          &Missed non-zero (\%) &N/A& \textbf{12.0}&1.15 &\textbf{0} &\textbf{0}  \\
    
    &Missed zero (\%) &N/A& 3.40& 1.22& 0.60&  0.17\\
       
          \hline
        \multirow{4}{4em}{$3.1623$}&$\mQ$&hub and spoke&0.569&0.492&0.492& 0.496 \\
        &$\mQ^{-1}$&block-diag& \textbf{0.819}& 0.796& 0.804&   0.812\\
          &Missed non-zero (\%) &N/A&  44.7& 45.9&49.3 & 55.0\\
    
    &Missed zero (\%) &N/A&   0.24& 0.03& 0.01 &\textbf{0}\\
        
         \hline
         
        \multirow{4}{4em}{$10$}&$\mQ$&hub and spoke&0.585&0.499&0.495&0.498  \\
        &$\mQ^{-1}$&block-diag&0.85 & 0.828& 0.824&   0.820\\
          &Missed non-zero (\%) &N/A& 65.3& 65.3& 65.3& 65.3 \\
    
    &Missed zero (\%) &N/A& \textbf{0}& \textbf{0}&\textbf{0} & \textbf{0} \\
        \hline
    \end{tabular}
    \caption{Median statistics for error in recovery of $\mQ$ and the sparsity pattern of $\mQ$ for varying $\lambda$ under a hub and spoke model. Median over 50 independent trials. Note that the parameters of the process $Z_2$ are not included here, since they are inferred before $\mQ$ and independent of $\lambda$. Best estimates for each value of $N$ are bolded.}
    \label{tab:med_errs_lambda_hubandspoke}
\end{table}

\newpage
\subsection{Random graph model for $\mQ$}
\begin{figure}[h]
    \centering
    \includegraphics[width=0.85\linewidth]{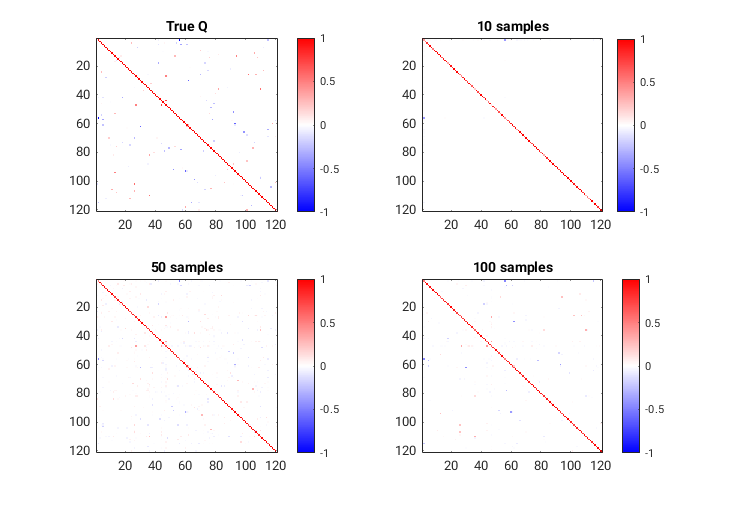}
    \caption{True and estimated $\mQ$ for the hub and spoke model. The true $\mQ$ is in the top left, and the remaining figures show the recovered $\mQ$ with 10, 50, and 100 samples of the field.}
    \label{fig:true_and_est_Q_randQ}
\end{figure}

\begin{table}[h!]
    \centering
    \begin{tabular}{|c|c|c|c|c|c|}
    \hline
       Parameter&Truth & N=10 &N=30  &N=50& N=100  \\
          \hline
    $\mQ$ & Random & 0.698 &0.422 &0.316 & 0.275\\      
          \hline
          $\mQ^{-1}$&Random& 0.471&0.139 &0.160 &  0.076 \\ \hline
           Missed non-zero (\%) &NA&52.8&49.1 &32.7&33.5 \\
    \hline
    Missed zero (\%)&NA &0.877& 1.61&6.1& 2.57\\
    \hline
    $\sigma^2$&1&0.241 & 0.120&0.101 &0.601\\
    \hline
    $\log_{10}(\tau^2)$&-2& 0.284&0.226& 0.210&0.200\\
    \hline
    $\theta$&0.3&0.213& 0.060& 0.055& 0.108\\
    \hline
    $r$& 0.15&0.075& 0.048& 0.052& 0.044\\
    \hline
    $\nu$& 0.5&0.086& 0.070& 0.054& 0.101\\
    \hline
    $L(\hat{\mQ},\hat{p})/L(\mQ,p)$ &1&0.976&0.990&1.04&0.96\\
    \hline
    \end{tabular}
    \caption{Median error in recovery of residual parameters for varying number of independent replicates of the field using a random graph model for $\mQ$, along with true values. All calculations are the median over 50 trials for a given number of independent replicates of the field. The error in the recovery of $\mQ$ is given by $\|\hat{\mQ}-\mQ\|/\|\mQ\|$ with $\|\cdot\|$ the Frobenius norm, and likewise for $\mQ^{-1}$, and the error for the other parameters is the absolute error. Notice that, while the FSBGL has the most trouble recovering $\mQ$ out of the simulated data cases presented here, it is able to capture $\mQ^{-1}$ far more accurately than in any other cases. }
    \label{tab:simresults_randq_appendix}
\end{table}

\begin{table}[h!]
    \centering
    \begin{tabular}{|c|c|c|c|c|c|c|}
    \hline
         $\lambda$&Parameter&Truth& N=10&N=30&N=50&N=100  \\
         \hline
        \multirow{4}{4em}{$0.1$}&$\mQ$&Random&5.06&1.62&0.893& 0.430\\
        &$\mQ^{-1}$&Random&0.476 & 0.141&0.161 & 0.076  \\
          &Missed non-zero (\%) &N/A& \textbf{35.7}&\textbf{21.9}&\textbf{17.8}& \textbf{14.9}\\
    
    &Missed zero (\%) &N/A& 19.4&34.1&35.6&29.8 \\
        \hline
           
        \multirow{4}{4em}{$0.31623$}&$\mQ$&Random&1.46&0.446&\textbf{0.315}&\textbf{0.275}  \\
        &$\mQ^{-1}$&Random& 0.475& 0.140& 0.160&  0.076 \\
          &Missed non-zero (\%) &N/A&39.4 &32.7&32.7&33.5 \\
    
    &Missed zero (\%) &N/A&13.7 &10.6&6.13& 2.57 \\
         \hline
        \multirow{4}{4em}{$1$}&$\mQ$&Random&0.827&\textbf{0.422}&0.366& 0.347 \\
        &$\mQ^{-1}$&Random&0.474 & 0.139& 0.160&   0.074\\
          &Missed non-zero (\%) &N/A& 49.1&49.1 &49.8 & 49.8 \\
    
    &Missed zero (\%) &N/A& 2.57& 1.61& 1.54&  1.4\\
       
          \hline
        \multirow{4}{4em}{$3.1623$}&$\mQ$&Random&0.777&0.436&0.387& 0.372 \\
        &$\mQ^{-1}$&Random& \textbf{0.463}&0.1355 & 0.160&0.071   \\
          &Missed non-zero (\%) &N/A&  52.0& 52.0&52.0 & 52.0\\
    
    &Missed zero (\%) &N/A&   1.43& 1.17& 1.0 &0.79\\
        
         \hline
         
        \multirow{4}{4em}{$10$}&$\mQ$&Random&\textbf{0.698}&0.449&0.410&  0.392\\
        &$\mQ^{-1}$&Random&0.471 & \textbf{0.130}& \textbf{0.159}&  \textbf{ 0.066}\\
          &Missed non-zero (\%) &N/A&52.8 &52.8&52.8&52.8\\
    
    &Missed zero (\%) &N/A& \textbf{0.88}& \textbf{0.42}&\textbf{0.22} & \textbf{0.10} \\
        \hline
    \end{tabular}
    \caption{Median statistics for error in recovery of $\mQ$ and the sparsity pattern of $\mQ$ for varying $\lambda$ under the random graph model. Median over 50 independent trials. Note that the parameters of the process $Z_2$ are not included here, since they are inferred before $\mQ$ and independent of $\lambda$. Best estimates for each value of $N$ are bolded.}
    \label{tab:med_errs_lambda_hubandspoke}
\end{table}
\newpage

\subsection{Smaller scale variation}

In addition to the residual model explored this far, we also explored the ability to recover a residual model with smaller Mat\'ern scale parameter under the block-diagonal model for $\mQ$. This greatly limited our ability to recover the taper range, but that parameter is much less important in governing the dynamics of the model in this case, and we are still able to get very near the value of the likelihood at the true parameters. For this reason we do not believe that this is an issue. It is also worth noting that the median error in the estimate for $r$ cannot be any lower due to the resolution of the grid that we are using to do the annealing.

\begin{table}[h!]
    \centering
    \begin{tabular}{|c|c|c|c|c|c|}
    \hline
        Parameter&Truth & N=10 &N=30  &N=50& N=100  \\
          \hline
    $\mQ$ & Block-diagonal & 0.628 &0.445 &0.436 & 0.277\\      
          \hline
           $\mQ^{-1}$ & block-diag & 0.653 &0.517 &0.423 & 0.436\\ \hline
    Missed non-zero (\%) &NA&52.0& 13.8 &8.00& 0 \\
    \hline
    Missed zero (\%)&NA &0.223& 4.9&1.53& 19.4\\
    \hline
    $\sigma^2$&1&0.081 & 0.061&0.061 &0.061\\
    \hline
    $\log_{10}(\tau^2)$&-2& 0.278&0.21& 0.254&0.290\\
    \hline
    $\theta$&0.3&0.675& 0.730& 0.705& 0.699\\
    \hline
    $r$& 0.05&0.001& 0.001& 0.001& 0.001\\
    \hline
    $\nu$& 0.5&0.038& 0.039& 0.039& 0.036\\
    \hline
    $L(\hat{\mQ},\hat{p})/L(\mQ,p)$ &1&1.00&0.983&0.992&0.983\\
    \hline
    \end{tabular}
    \caption{Median error in recovery of residual parameters with small-scale variation in the residuals using a block-diagonal model for $\mQ$, along with true values. All calculations are the median over 50 trials for a given number of independent replicates of the field. The error in the recovery of $\mQ$ is given by $\|\hat{\mQ}-\mQ\|/\|\mQ\|$ with $\|\cdot\|$ the Frobenius norm, and the error for the other parameters is the absolute error.}
    \label{tab:simresults_smallscale_appendix}
\end{table}
\clearpage
\section{Basis Comparison}\label{sec:basis_comp}

In Section \ref{sec:extant_comp} and elsewhere we make the claim that the choice of basis does not drive the change in performance between the models, rather it is due to the inclusion of the small scale structure. To demonstrate this, we also ran the FSBGL using a Wendland basis and a tapered Mat\'ern covariance model to examine the effect on performance and recovered small-scale parameters. Those results are presented here. 

First, we examine the retrieved parameters of the small-scale covariance model. The basis expression enters the estimation via the assumption that the coefficients are IID, and a diagonal covariance is fitted jointly with the small scale parameters (Eq. \eqref{eq:invertD}). The only parameter that sees meaningful change is the nugget variance, and that is still only about $0.58$ vs $0.15$. Additionally, both are orders of magnitude smaller than the nugget variances fit by the other models (see Table \ref{tab:params}). 

\begin{table}[h!]
    \centering
    \begin{tabular}{c|c|c|c|c|c}
       Model & $\tau^2$ &$\sigma^2$ &Support&Smoothness&Range\\
        \hline
        Needlets+TM & $0.15$ &$1032$&$27.6\degrees$&$0.63$&$7.5\degrees$\\
        Wendland+TM& $0.58$ &$968$& $26.9\degrees$&$0.62$&$7.2\degrees$ \\
        Percent change & $286$& $6.2$ &$2.5$ &$1.6$ &$4.0$ \\
    \end{tabular}
    \caption{Fitted tapered Mat\'ern parameters for $Z_2$ using both a needlet and Wendland basis, along with their percent differences. The support parameter is the angular half-width of the support of the covariance function.}
    \label{tab:params_needvwend}
\end{table}

The results related to predictive performance are included in Table \ref{tab:predictions_basiscomp}, as well as the negative log-likelihoods on the test data. Their predictive performances are effectively identical, as are the log-likelihoods. This provides additional support for our assertion that the main source of improvement in the performance of the model is the addition of the small-scale structure rather than the choice of basis.  

\begin{table}[h!]
    \centering
    \begin{tabular}{c|c|c|c|c}
         Model& Negative log-likelihood (test set) &Mean CRPS & Median CRPS & RMSE \\
         \hline
         Needlets+TM&9.6$\times 10^4$&5.58&3.15&10.51\\
         Wendland+TM&9.4$\times 10^4$&5.56&3.16&10.49\\
         Percent change&2.0&0.4&0.3&0.2
    \end{tabular}
    \caption{Mean and median CRPS, as well as RMSE, for subgrid scale predictions from the FSBGL using a needlet and Wendland basis, along with the percent change. Both models used the tapered Mat\'ern residual model with parameters in Table \ref{tab:params_needvwend}.}
    \label{tab:predictions_basiscomp}
\end{table}

\clearpage
\section{Sample Autocorrelation of the 
WACCM-X Data} \label{sec:autocorr_appendix}

Due to the fact that our samples are output from a time-dependent model, we want to verify that there is limited sample autocorrelation in our data. For this reason, we computed the autocorrelation functions at the spatial locations across samples, a subset of which are plotted here with the corresponding latitude and solar local time. As we can see, the sample autocorrelation is generally small, so we feel comfortable treating these samples as independent.

\begin{figure}[h!]
    \centering
    \includegraphics[width=\linewidth]{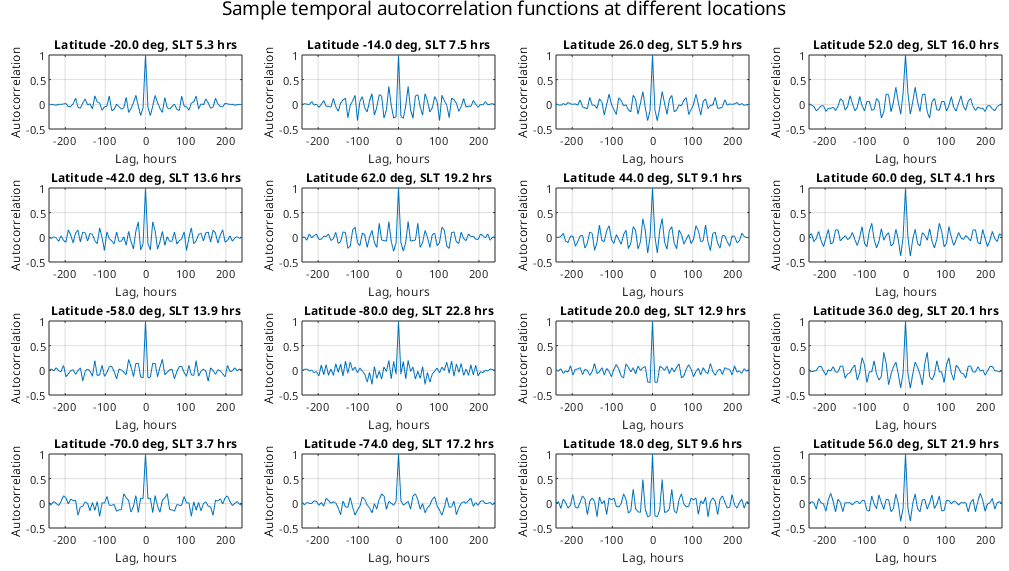}
    \caption{Sample autocorrelation functions at several latitudes and solar local times (SLT)}
    \label{fig:autocorr}
\end{figure}

\clearpage
\section{Behavior With Changes in $\mD$}\label{sec:nuggeffect}

In Section \ref{sec:modelselection} we hypothesize that the reason that the nugget effect is much smaller for the Tapered Mat\'ern model is that the parameter $\nu$ allows the model to fit the smoothness of the field, while the Gaspar-Cohn and Wendland models do not have the same capability. 
We provide in this appendix some empirical validation of that assertion. 

To investigate the behavior further, we generated 10, 30, 50, and 100 samples of a field on $[0,1]^2$ using the same cosine basis as the other simulation studies and a block-diagonal precision matrix model. The residual field has the same properties as in \ref{sec:simdata_appendix_bd}, that is, a tapered Mat\'ern covariance with parameters $\tau^2=0.01,\sigma^2=1,\theta=0.3,r=0.15, \nu=0.5$. However, we are examining whether the Wendland covariance model predicts a larger nugget effect. We believe that since the Mat\'ern smoothness is small, leading to a field that is mean-square continuous but not differentiable, and the Wendland covariance model produces smooth fields, we ought to see a much larger nugget effect variance fit in order to compensate. The results of our experiment are shown in Table \ref{tab:nuggetstuff}. We see that across sample sizes the estimates of the nugget effect using the Wendland model are greatly inflated, with the fitted nugget standard deviation being 7 times larger under the Wendland model. This supports our expectation that the other models may attempt to fit the rougher fields by exaggerating the nugget effect. 

\begin{table}[h!]
\centering
      \begin{tabular}{|c|c|c|c|c|c|}
    \hline
        Model&Truth & N=10 &N=30  &N=50&N=100  \\
          \hline
    Tapered Mat\'ern& -2&-2.07&-2.05& -2.01&-2.02\\
          \hline
    Wendland covariance &-2&-0.302& -0.301 &-0.306& -0.304 \\
    \hline
    \end{tabular}
     \caption{Mean estimate of $\log_{10}(\tau^2)$ on a Mat\'ern residual field with $\tau^2=0.01$, for both the Mat\'ern model and a Wendland model. The Wendland model selects a much larger nugget effect, supporting our belief that this might be related to the smoothness of the underlying field. }\label{tab:nuggetstuff}
    \end{table}


\newpage

\bibliographystyle{elsarticle-harv} 
\bibliography{references}


\end{document}